\documentstyle[11pt,axodraw]{article}
\begin{document}
\setcounter{footnote}{0}
\renewcommand{\thefootnote}{\alph{footnote}}
\renewcommand{\theequation}{\thesection.\arabic{equation}}
\newcounter{saveeqn}
\newcommand{\add}{\addtocounter{equation}{1}}
\newcommand{\alpheqn}{\setcounter{saveeqn}{\value{equation}}%
\setcounter{equation}{0}%
\renewcommand{\theequation}{\mbox{\thesection.\arabic{saveeqn}{\alph{equation}}}}}
\newcommand{\reseteqn}{\setcounter{equation}{\value{saveeqn}}%
\renewcommand{\theequation}{\thesection.\arabic{equation}}}
\newenvironment{nedalph}{\add\alpheqn\begin{eqnarray}}{\end{eqnarray}\reseteqn}
\newsavebox{\PSLASH}
\sbox{\PSLASH}{$p$\hspace{-1.8mm}/}
\newcommand{\PS}{\usebox{\PSLASH}}
\newsavebox{\PARTIALSLASH}
\sbox{\PARTIALSLASH}{$\partial$\hspace{-2.3mm}/}
\newcommand{\PARTIALS}{\usebox{\PARTIALSLASH}}
\newsavebox{\ASLASH}
\sbox{\ASLASH}{$A$\hspace{-2.1mm}/}
\newcommand{\AS}{\usebox{\ASLASH}}
\newsavebox{\QSLASH}
\sbox{\QSLASH}{$q$\hspace{-2.1mm}/}
\newcommand{\QS}{\usebox{\QSLASH}}
\newsavebox{\KSLASH}
\sbox{\KSLASH}{$k$\hspace{-1.8mm}/}
\newcommand{\KS}{\usebox{\KSLASH}}
\newsavebox{\LSLASH}
\sbox{\LSLASH}{$\ell$\hspace{-1.8mm}/}
\newcommand{\LS}{\usebox{\LSLASH}}
\newsavebox{\LLSLASH}
\sbox{\LLSLASH}{$L$\hspace{-1.8mm}/}
\newcommand{\LLS}{\usebox{\LLSLASH}}
\newsavebox{\SSLASH}
\sbox{\SSLASH}{$s$\hspace{-1.8mm}/}
\newcommand{\SS}{\usebox{\SSLASH}}
\newsavebox{\DSLASH}
\sbox{\DSLASH}{$D$\hspace{-2.8mm}/}
\newcommand{\DS}{\usebox{\DSLASH}}
\newsavebox{\DbfSLASH}
\sbox{\DbfSLASH}{${\mathbf D}$\hspace{-2.8mm}/}
\newcommand{\DBFS}{\usebox{\DbfSLASH}}
\newsavebox{\DELVECRIGHT}
\sbox{\DELVECRIGHT}{$\stackrel{\rightarrow}{\partial}$}
\newcommand{\PARVECR}{\usebox{\DELVECRIGHT}}
\thispagestyle{empty}
\begin{flushright}
IPM/P-2002/013
\par
hep-th/0206009
\end{flushright}
\vspace{0.5cm}
\begin{center}
{\Large\bf{Noncommutative Dipole QED}}\\
\vspace{1cm} {\bf N\'eda Sadooghi$\
^{\dagger,}$}\footnote{\normalsize{Electronic address:
sadooghi@theory.ipm.ac.ir}} \hspace{0.2cm} and \hspace{0.2cm}{\bf
Masoud Soroush$\ ^{\ddagger,}$}\footnote{\normalsize{Electronic
address:
soroush@mehr.sharif.ac.ir}}  \\
\vspace{0.5cm}
{\sl ${\ ^{\dagger, \ddagger}}$ Department of Physics, Sharif University of Technology}\\
{\sl P.O. Box 11365-9161, Tehran-Iran}\\
and\\
{\sl ${\ ^{\dagger}}$ Institute for Studies in Theoretical Physics and Mathematics (IPM)}\\
{\sl{School of Physics, P.O. Box 19395-5531, Tehran-Iran}}\\
\end{center}
\vspace{0cm}
\begin{center}
{\bf {Abstract}}
\end{center}
\begin{quote}
The noncommutative dipole QED is studied in detail for the matter
fields in the adjoint representation. The axial anomaly of this
theory is calculated in two and four dimensions using various
regularization methods. The Ward-Takahashi identity is proved by
making use of a non-perturbative path integral method. The
one-loop $\beta$-function of the theory is calculated explicitly.
It turns out that the value of the $\beta$-function depends on the
direction of the dipole length $\vec{L}$, which defines the
noncommutativity. Finally using a semi-classical approximation a
non-perturbative definition of the form factors is presented and
the anomalous magnetic moment of this theory at one-loop order is
computed.
\end{quote}

\hspace{0.8cm}
\newpage
\setcounter{page}{1}
\section{Introduction}
In the past few years, the noncommutative theories are studied
intensively by many authors. Especially Moyal noncommutative gauge
theories are interesting due to their realization in the String
Theory. As it turns out, in the decoupling limit, noncommutative
gauge theories can occur in the world volume of $Dp$-branes in the
presence of constant background $B_{\mu\nu}$ field, with $\mu$ and
$\nu$ lying both on the branes \cite{h0}. For a review of Moyal
noncommutativity see Refs. \cite{h1, h2}.
\par
In Refs. \cite{h3, h4}, however, a new type a noncommutative
product is introduced by considering a system of $Dp$-branes in
the presence of a constant background $B_{\mu\nu}$ field with one
index along the branes' world volume and the other index
transverse to it. The gauge theories defined on these $Dp$-branes
are also noncommutative, but, in contrast to the Moyal case, the
space-time remains commutative. The noncommutativity appears only
in the product of functions and has its origin in the finite
dipole length $\vec{L}$ associated to each field. A supergravity
description of these so called noncommutative dipole theories is
presented in \cite{h5}. As in the Moyal case, the noncommutative
dipole Field Theory can also be defined by replacing the ordinary
product of function by a noncommutative dipole $\star$-product.
The noncommutative dipole Field Theory is studied first in Ref.
\cite{h6}.
\par
In this paper, a detailed study of the noncommutative dipole QED
is presented. In this theory, in analogy to the Moyal case, the
matter fields appear in fundamental, antifundamental and adjoint
representations. Here, we will restrict ourselves to
noncommutative dipole QED with matter fields in the adjoint
representation. In this theory point-like charged particles are
absent. The only interacting objects are multipoles. Hence
noncommutative dipole QED with adjoint matter fields is an
appropriate candidate to study the interaction of neutral
particles with finite dipole moments, like neutrinos, with gauge
particles like photons. There are some experimental evidences of
such interactions, which cannot be described by the commutative
version of the standard model of particles \cite{h7}. This is not
the only ground to study this adjoint theory. In the framework of
perturbative calculations planar as well as nonplanar Feynman
diagrams appear, which make the theory non-trivial. As in the
noncommutative Moyal gauge theory, UV/IR mixing effects \cite{h9}
can appear for small dipole length and large momentum cutoff.
\par
In Sect. 2, a brief description of the algebraic structure of
noncommutative dipole Field Theory is presented. The action of the
noncommutative dipole QED in the adjoint representation is
introduced in Sect. 3, where its global symmetries are also
studied. As in the noncommutative Moyal case \cite{h10, h11}, the
theory possesses three different currents. We will show that only
two of them correspond to finite conserved axial charges. In Sect.
4, the axial anomalies of these currents are calculated in two and
four dimensions using point split and dimensional regularization
methods. In Ref. \cite{h12}, the axial anomaly arising from only
one of the currents of the theory is calculated using the
Fujikawa's path integral method \cite{h14}. Our results coincides
with the results presented in this paper.
\par
In Sect. 5, a one-loop perturbative analysis is carried out to
compute the one-loop contributions to the fermion-self energy,
vacuum polarization tensor and the vertex function. We will show
that one-loop fermion-self energy and vertex function include
planar and nonplanar parts, whereas the one-loop vacuum
polarization tensor exhibits only a planar Feynman integral. We
will calculate the one-loop contributions to the renormalization
constants $Z_{i}, i=1,2,3$, and show that $Z_{1}=Z_{2}$ and that
$Z_{3}$ is proportional to the tensorial structure
$(p_{\mu}p_{\nu}-p^{2}\eta_{\mu\nu})$. To show that theses
identities are also valid in all higher orders of perturbative
expansion, we will prove the Ward-Takahashi identity of the
noncommutative dipole QED in the adjoint representation in Sect.
6. This will be done first in the framework of perturbation theory
and then using the non-perturbative path integral method. In Sect.
7, the Ward-Takahashi identity will be used to obtain the general
forms for the renormalization constant $Z_{i}, i=1,2,3$.
\par
We will then calculate explicitly the one-loop $\beta$-function of
noncommutative QED with adjoint matter fields [see Sect. 7.3]. It
is shown that, in contrast to the commutative QED, this theory is
asymptotically free, and that the one-loop $\beta$-function is
proportional to a non-negative factor $(\vec{p}\cdot
\vec{L})^{2}\equiv |\vec{p}||\vec{L}|\cos\vartheta$, where
$\vec{p}$ is a small external momentum, $\vec{L}$ is the dipole
length associated to each matter fields, and $\vartheta$ is the
relative angle between these two vectors. The value of the
$\beta$-function depends therefore on the lengths of two vectors
$\vec{p}$ and $\vec{L}$, as well as on the relative direction of
$\vec{L}$ and $\vec{p}$.
\par
The factor $(\vec{p}\cdot \vec{L})^{2}$ appearing in the one-loop
$\beta$ function of the theory, has in fact an interesting
physical origin: As is known, for small external momentum the
scattering amplitudes has to coincide with the classical results.
According to the Born approximation, the scattering amplitude is
proportional to the Fourier transformed of the scattering
potential energy. In a theory where point-like charged particles
exist, like in ordinary commutative QED, this potential energy is
the Coulomb potential. But in our noncommutative dipole QED in the
adjoint representation, this potential energy is the energy
between multipoles. We have found that the factor $(\vec{p}\cdot
\vec{L})$, which appears in the one-loop $\beta$-function of the
theory, arises in fact from the Fourier transformed of the
potential energy between two dipoles with the dipole moments
$g_{0}\vec{L}$ (see Appendix A for a derivation), defining a
modified bare coupling constant $\bar{g}_{0}=(\vec{p}\cdot
\vec{L})g_{0}$.
\par
Using a semi-classical approximation in Sect. 8, we have shown
that the form factors of noncommutative dipole QED with adjoint
matter fields, can be defined by the Fourier transformed of the
potential energy between a {\it dipole} $g\vec{L}$ and external
electric and magnetic potentials. The anomalous magnetic moment is
then calculated in one-loop order. Sect. 9 is devoted to
discussions.
\section{Algebraic Structure of Dipole Field Theory}
Let us establish a noncommutative Field Theory, by defining  a
$\star$-product in the linear space ${\mathcal{A}}$ of quantum
fields and forming an associative complex $C^{\star}$ algebra with
respect to this product. A noncommutative space is then defined by
an automorphism ${\mathcal{Q}}: {\mathcal{A}}\to {\mathcal{A}}$
which defines a  derivative on the algebra ${\mathcal{A}}$ and a
linear map from ${\mathcal{A}}$ to the field $C$. This map is then
given by $\int: {\mathcal{A}}\to C$ which acts as a trace on this
algebra. If the following three properties
\begin{enumerate}
\item the Leibnitz Rule: $\forall\  \Phi_{a}, \Phi_{b}\in {\mathcal{A}}$
\begin{eqnarray}\label{S11}
{\mathcal{Q}}(\Phi_a \star
\Phi_b)=({\mathcal{Q}}\Phi_a)\star\Phi_b+\Phi_a\star({\mathcal{Q}}\Phi_b),
\end{eqnarray}
\item integration by part, {\it i.e.} $\forall\  \Phi_{a}\in {\mathcal{A}}$
\begin{eqnarray}\label{S12}
\int {\mathcal{Q}}(\Phi_a) =0,
\end{eqnarray}
\item and the cyclicity, {\it i.e.} $\forall\  \Phi_{a}, \Phi_{b}\in {\mathcal{A}}$
\begin{eqnarray}\label{S13}
\int (\Phi_a\star\Phi_b)=\int (\Phi_b\star\Phi_a),
\end{eqnarray}
\end{enumerate}
are satisfied for the above two maps, then the collection
$({\mathcal{A}}, {\mathcal{Q}},\int)$ forms a noncommutative space
(for a  review see \cite{h1}). The noncommutative Field Theory is
then defined on this noncommutative space and its action is built
using the above maps ${\mathcal{Q}}$ and $\int$. To construct a
noncommutative dipole Field
  Theory a constant dipole $L_{a}^{\mu}=(0,L_{a}^{i})$ is assigned to
each element $\Phi_{a}$
 of the algebra. The dipole $\star$-product is then defined as follows:
\begin{eqnarray}\label{S14}
 \star: && C^\infty({\mathcal{R}}^4)\otimes C^\infty({\mathcal{R}}^4)\longrightarrow
 C^\infty({\mathcal{R}}^4)\nonumber\\
(\Phi_a\star\Phi_b)(x)&\equiv&\Phi_a(x-L_b/2) \Phi_b(x+L_a/2).
\end{eqnarray}
The dipole length corresponding to the product
$\Phi_{a}\star\Phi_{b}$ is, due to the associativity
 of the algebra, given by the sum of the dipoles of $\Phi_{a}$ and $\Phi_{b}$ \cite{h6}.
  The derivative on the noncommutative dipole space is defined by the ordinary derivative
   of functions and satisfies automatically the Leibnitz rule.
\par
As in the Moyal case, the trace on the algebra is given by the
integration over all space-time components, so that the second
property [Eq. \ref{S12})] of the noncommutative space is easily
satisfied. The third property (\ref{S13}), however, is satisfied
only for the kernel of the following map $S_{n}$:
\begin{eqnarray}\label{S15}
S_{n}: \bigotimes_{i=1}^{n}C^{\infty}({\mathcal{R}}^{4})\to
{\mathcal{R}}^{4},\hspace{1cm}
S_{n}(\Phi_{1},\Phi_{2},\cdots,\Phi_{n})=\sum_{i=1}^{n}L_{i}.
\end{eqnarray}
For physical purposes, we restrict ourselves to this
kernel\footnote{As is known, in noncommutative dipole Field
Theory, the space-time coordinates are still commutative. The
space and time are therefore homogeneous and this means that the
Lagrangian densities of QFTs are, in general, translational
invariant. As a consequence, in addition to a energy-momentum
conservation, the sum of the dipole lengths associated to each
term in the Langrangian must vanish.}. For the fields $\Phi$
belonging to the $C^{\star}$-algebra, the product
$(\Phi\star\Phi^{\dagger})(x)$ is real valued. The dipole length
assigned to the self-adjoint of a field is therefore given by the
negative value of the dipole length associated to the field
itself. As  a consequence, the dipole length corresponding to
Hermitean fields is zero.
\par
If the sum of the dipole lengths of two fields $\Phi_{a}$ and
$\Phi_{b}$ vanishes, {\it i.e.} if $(\Phi_{a}, \Phi_{b}) \in
\mbox{Ker}\ S_{2} $, we have:
\begin{eqnarray}\label{S16}
\int d^{4}x\ (\Phi_{a}\star\Phi_{b})(x)=\int d^{4}x \
\Phi_{a}(x)\Phi_{b}(x).
\end{eqnarray}
This can be shown by a simple change of integration variable.
Note that the same property is also valid for the Moyal
$\star$-product, only if both fields $\Phi_{a}$ and $\Phi_{b}$
have trivial boundary conditions.
\section{Dipole QED in the Adjoint Representation}
\setcounter{equation}{0}
 Consider the action of noncommutative
dipole QED, that includes the pure gauge part and the matter field
part:
\begin{eqnarray}\label{S17}
S_{QED}[\psi,\bar{\psi},A]= \int\  F_{\mu\nu}\star F^{\mu\nu}+
  \int\  \bar{\psi}\star(i\DS-m)\psi.
\end{eqnarray}
Here, the field strength tensor is defined as in the ordinary
commutative QED by $F_{\mu\nu}=\partial_{[\mu}A_{\nu]}$. Note that
in the Moyal case, the field strength tensor includes a Moyal
bracket of two gauge fields $A_{\mu}$. This nonlinear term leads
to new three and four gauge vertices. As it turns out these
vertices are absent in the present noncommutative dipole QED.
\par
The matter field part includes the covariant derivative, defined
by:
\begin{eqnarray}\label{S18}
D_{\mu}\equiv \partial_{\mu}+ig A_{\mu}.
\end{eqnarray}
Due to the noncommutativity of the dipole $\star$-product, the
matter fields have, in analogy to the Moyal case,  three different
representations: the fundamental, antifundamental and the adjoint
representations \cite{h6}. The action of the fermions in the
adjoint representation is given by:
\begin{eqnarray}\label{S19}
S^{adj}_{QED}[\psi,\bar{\psi},{A}]=\int
   F_{\mu\nu}\star F^{\mu\nu}+\int\
   \bar{\psi}\star(i\PARTIALS-m)\psi
   -g\int\
   \bar{\psi}\gamma^{\mu}\star[A_{\mu},\psi]_{\star}.
\end{eqnarray}
The action ({\ref{S19}) is invariant under the following
transformation of matter and gauge fields:
\begin{eqnarray}\label{S20}
\psi&\longrightarrow&(U\star\psi\star U^{-1}),\hspace{1cm}U\in
\textsl{U}(1)\nonumber\\
A&\longrightarrow&
  U\star\textsl{A}\star U^{-1}+\frac{i}{g}(\partial
  U\star U^{-1})=\textsl{A}+\frac{i}{g}(\partial U)U^{-1}.
\end{eqnarray}
The dipole $\star$-product in the expression
$U\star\textsl{A}\star U^{-1}$ could be removed, because both
$A_{\mu}$ and $U(x)$ are dipoleless. Using Eq. (\ref{S16}), the
dipole $\star$-product can be removed from the free part of the
action (\ref{S19}), too. Hence, the fermion and photon propagators
are exactly the same as in the ordinary commutative QED. The
vertex of two fermions and one gauge field must, however, be
modified and is given by:
 \vskip0.2cm
\begin{eqnarray}\label{S21}
\SetScale{1}
    \begin{picture}(50,20)(0,0)
    \Vertex(0,0){2}
    \Photon(0,0)(0,20){2}{4}
    \LongArrow(5,18)(5,12)
    \ArrowLine(-20,-20)(0,0)
    \ArrowLine(20,-20)(0,0)
    \Text(20,15)[]{$k,\mu$}
    \Text(-25,-15)[]{$p_{1}$}
    \Text(25,-15)[]{$p_{2}$}
    \end{picture}
\hspace{0.5cm}
V_{\mu}\left(p_{1},p_{2};k\right)=-\left(2\pi\right)^{4}\delta^{4}
\left(p_{1}+p_{2}+k\right)2g\gamma_{\mu}\ \sin\left(\frac{k\cdot
L}{2}\right).
\end{eqnarray}
\vspace{0.3cm}
\subsection{Noether Currents and Conserved Charges}
In this section the Noether currents and the corresponding
conserved charges of the noncommutative dipole U(1) gauge theory
with adjoint matter fields will be derived explicitly.
\par
Consider first the action $S^{adj.}[\psi,\bar{\psi},A]$ of a
noncommutative Field Theory with adjoint matters, which is
invariant under an arbitrary global and continuous symmetry
transformation of matter fields
\begin{eqnarray}\label{S22}
\psi\to
\psi+\epsilon{\mathcal{F}}(\psi),\hspace{1cm}\mbox{and}\hspace{1cm}\bar{\psi}\to
\bar{\psi}+\epsilon{\mathcal{F}}^{*}(\bar{\psi}).
\end{eqnarray}
Here, $\epsilon$ is a constant real valued number. The variation
of the action under a local infinitesimal transformation is given
by:\footnote{This infinitesimal transformation is an arbitrary
one. It can be equivalently given by
$\psi\to\psi+\epsilon\star{\mathcal{F}}(\psi)$ or
\linebreak$\psi\to \psi+{\mathcal{F}}(\psi)\star\epsilon$.}
\begin{eqnarray}\label{S23}
 S^{adj.}\big[\psi+[\epsilon,{\mathcal{F}}(\psi)]_{\star},\bar{\psi}+
 [\epsilon,{\mathcal{F}}^{*}(\bar{\psi})]_{\star},A\big]-S^{adj.}\big[
 \psi,\bar{\psi},A\big]=
 -\int{d^{4}x\ J^{\mu}\big(\psi(x),\bar{\psi}(x)\big)\partial_{\mu}\epsilon(x)},
\end{eqnarray}
where $\epsilon(x)\in C^{\infty}({\mathcal{R}}^{4})$. For the
classical path in which the fields satisfy the equation of motion,
the r.h.s. of the above equation vanishes for all $\epsilon(x)$.
According to the property (\ref{S16}), the most general form for
the divergence of the current is given by:
\begin{eqnarray}\label{S24}
\partial_{\mu}J^{\mu}=\big[f(\psi,\bar{\psi}),g(\psi,\bar{\psi})\big]_{\star},
\end{eqnarray}
where $f$ and $g$ are arbitrary functions of $\psi$ and
$\bar{\psi}$. To find the conserved charge corresponding to the
current $J_{\mu}$, let us consider the three dimensional volume
integral over the divergence of the current. Using the equation
(\ref{S24}), the continuity equation is given by:
\begin{eqnarray}\label{S25}
\frac{dQ}{dt}-\int d^{3}x\ \partial_{i}J_{i}(x)=\int d^{3}x\
\big[f(\psi,\bar{\psi}),g(\psi,\bar{\psi})\big]_{\star},
\end{eqnarray}
where the charge $Q$ is defined by $Q\equiv \int d^{3}x J^{0}(x)$,
as in the ordinary commutative Field Theory. Since the time
component of the dipole length corresponding to the fermionic
fields $\psi$ and $\bar{\psi}$ is defined to be zero, the r.h.s.
of the Eq. (\ref{S25}) vanishes, defining a conserved charge.
\par
This general formulation can be used to find the local vector
current of the noncommutative dipole QED with adjoint matter
fields. The action (\ref{S19}) is invariant under the following
global transformation of matter fields:
\begin{eqnarray}\label{S26}
 \psi\longrightarrow e^{i\alpha}\psi,\quad\quad\bar{\psi}\longrightarrow \bar{\psi}
 e^{-i\alpha},
 \end{eqnarray}
where $\alpha$ is a real valued number. Considering now the
Langrangian density corresponding to this action:
\begin{eqnarray}\label{S27}
 {\mathcal{L}}^{adj.}_{QED}(\psi,\bar{\psi},A)=\bar{\psi}\star(i\PARTIALS-m)\psi-
 \frac{1}{4}\big(F_{\mu\nu}\big)^{2}
 -g\bar{\psi}\gamma^{\mu}\star[A_{\mu},\psi]_{\star}.
\end{eqnarray}
and a local infinitesimal transformation:
\begin{eqnarray}\label{S28}
 \psi\longrightarrow(1+i\alpha)\star\psi\star(1-i\alpha)\simeq\psi
 +i[\alpha,\psi]_{\star},
\end{eqnarray}
the Lagrangian density transforms as:
\begin{eqnarray}\label{S29}
 {\mathcal{L}}^{adj.}_{QED}\longrightarrow{\mathcal{L}}^{adj.}_{QED}-
 g\bar{\psi}\gamma^{\mu}\star[\partial_{\mu}\alpha,\psi]_{\star}.
\end{eqnarray}
Going through the same formulation described before, the vector
current of this theory can be given by:
\begin{eqnarray}\label{S30}
 J_{\mu}(x)=-g(\gamma_{\mu})^{\alpha\beta}
 \big\{\psi_{\beta},\bar{\psi}_{\alpha}\big\}_{\star}(x).
\end{eqnarray}
As in the Moyal case, this theory possesses two other global
vector currents \cite{h11}:
\begin{eqnarray}\label{S31}
 J'_{\mu}(x)&=&-g(\gamma_{\mu})^{\alpha\beta}(\psi_
{\beta}\star\bar{\psi}_{\alpha})(x)\nonumber\\
 J''_{\mu}(x)&=&-g(\gamma_{\mu})^{\alpha\beta}(\bar{\psi}_
{\alpha}\star\psi_{\beta})(x),
\end{eqnarray}
which correspond to the local infinitesimal transformations
$\psi\to (1+i\alpha)\star\psi$ and $\psi\to \psi\star(1+i\alpha)$,
respectively. Using the equation of motion corresponding to the
Lagrangian density (\ref{S28}), it can be easily shown that all
these currents are classically conserved:
\begin{eqnarray*}
\partial^{\mu}{\mathcal{J}}_{\mu}=0,\hspace{1cm}\mbox{for}\hspace{1cm}
{\mathcal{J}}_{\mu}\in \{J_{\mu},J'_{\mu},J''_{\mu}\}.
\end{eqnarray*}
\section{Axial Anomaly} \setcounter{equation}{0}
In this section we will study the axial anomaly of noncommutative
dipole QED with matter fields in the adjoint representation. First
we study the axial anomaly corresponding to the global axial
vector current
\begin{eqnarray}\label{S32}
J_{\mu(5)}(x)=-(\gamma_{\mu}\gamma_{5})^{\alpha\beta}
\big\{\psi_{\beta},\bar{\psi}_{\alpha}\big\}_{\star}(x).
\end{eqnarray}
Here we will use two different regularization methods: the point
split and dimensional regularization  in two and four
dimensions\footnote{When this study was almost done an article
appeared \cite{h12}, where the axial anomaly of noncommutative
dipole QED corresponding to the axial vector current (\ref{S32})
was calculated using the Fujikawa's path integral method.}. We
then calculate the axial anomaly corresponding to the two other
global axial vector currents of the theory $J'_{\mu(5)}$ and
$J''_{\mu(5)}$:
\begin{eqnarray}\label{S33}
 J'_{\mu(5)}(x)&=&-(\gamma_{\mu}\gamma_{5})^{\alpha\beta}(\psi_
{\beta}\star\bar{\psi}_{\alpha})(x)\nonumber\\
 J''_{\mu(5)}(x)&=&-(\gamma_{\mu}\gamma_{5})^{\alpha\beta}(\bar{\psi}_
{\alpha}\star\psi_{\beta})(x),
\end{eqnarray}
using only dimensional regularization. \subsection{The Axial
Anomaly of $J_{\mu(5)}$}
\subsubsection{Point Splitting Regularization}
\par\noindent
{\it i) Two Dimensions}
\par\noindent
Consider the axial vector current $J_{\mu(5)}$ from Eq.
(\ref{S32}), whose point splitted version reads:
\begin{eqnarray}\label{S34}
 J_{\mu(5)}(x)&=&
 \mbox{symm}\lim_{\epsilon\longrightarrow0}(\gamma^{\mu}\gamma^{5})^{\alpha\beta}
 \Big[\psi_{\beta}(x-\frac{\epsilon}{2})\star {\mathcal{U}}^{\dag}(x+\frac{\epsilon}{2},
 x-\frac{\epsilon}{2})
 \star\bar{\psi}_{\alpha}(x+\frac{\epsilon}{2})\nonumber\\
 &&+\bar{\psi}_{\alpha}(x+\frac{\epsilon}{2})\star{\mathcal{U}}(x+\frac{\epsilon}{2},
 x-\frac{\epsilon}{2})
 \star\psi_{\beta}(x-\frac{\epsilon}{2})\Big].
\end{eqnarray}
Here, we have introduced the link variables:
\begin{eqnarray}\label{S35}
{\mathcal{U}}(y,x)\equiv\exp\Big(-ig\int_{x}^{y}dz^{\mu}A_{\mu}(z)\Big),
\end{eqnarray}
with the known gauge transformation property:
\begin{eqnarray}\label{S36}
{\mathcal{U}}(x,y)\to U(x)\star{\mathcal{U}}(x,y)\star
U^{\dagger}(y),
\end{eqnarray}
where $U\in U(1)$-gauge group.  Using the transformation
(\ref{S20}) of the matter fields and Eq. (\ref{S36}) for the link
variables, it can be shown that $J_{\mu(5)}(x)$ from Eq.
(\ref{S34}) is gauge invariant. This is because the product of
$\psi$, $\bar{\psi}$ and ${\mathcal{U}}$, appearing on the r.h.s.
of Eq. (\ref{S34}), is dipoleless. Similar calculation is also
performed in \cite{h10} for the Moyal case, where, however, the
regularized currents is only gauge {\it covariant}. An integration
over all (Moyal) noncommutative space-time coordinates has to be
performed to preserve the gauge invariance.
\par
After this remark, we use the expansion of the link variable in
the first order of $\epsilon$:
\begin{eqnarray}\label{S37}
{\mathcal{U}}(x+\frac{\epsilon}{2},x-\frac{\epsilon}{2})=
1-ig\epsilon^{\mu}A_{\mu}(x)+{\mathcal{O}}(\epsilon^{2}).
\end{eqnarray}
The VEV of the divergence of the axial vector current is then
given by:
\begin{eqnarray}\label{S38}
\lefteqn{ \Big<\partial^{\mu}J_{\mu(5)}(x)\Big>=\mbox{symm}
\lim_{\epsilon\longrightarrow0}g\epsilon^{\nu}
 (\gamma^{\mu}\gamma^{5})^{\alpha\beta}
 \Biggr(2\Big<\Big[\psi_{\beta}(x-\frac{\epsilon}{2}),\bar{\psi}_{\alpha}
 (x+\frac{\epsilon}{2})\Big]_{\star}\Big>\partial_{\nu}A_{\mu}(x)}\nonumber\\
 &&+\Big<\psi_{\beta}(x-\frac{\epsilon}{2})\star\bar{\psi}_{\alpha}
 (x+\frac{\epsilon}{2})\Big>F_{\mu\nu}(x+L)-
 \Big<\bar{\psi}_{\alpha}(x+\frac{\epsilon}{2})\star\psi_{\beta}
 (x-\frac{\epsilon}{2})\Big>F_{\mu\nu}(x-L)\Biggr).
\end{eqnarray}
In the zeroth order of perturbative expansion, using the
definition of dipole $\star$-product, the expectation value of
matter fields can be easily calculated:
\begin{eqnarray}\label{S39}
(\gamma^{\mu}\gamma^{5})^{\alpha\beta}
\Big<\psi_{\beta}(x-\frac{\epsilon}{2})\star\bar{\psi}_{\alpha}
 (x+\frac{\epsilon}{2})\Big>&=&
 (\gamma^{\mu}\gamma^{5})^{\alpha\beta}
 \Big<\psi_{\beta}(x+\frac{L}{2}-\frac{\epsilon}{2})
 \bar{\psi}_{\alpha}(x+\frac{L}{2}+\frac{\epsilon}{2})\Big>\nonumber\\
  &=&(\gamma^{\mu}\gamma^{5})^{\alpha\beta}\int{\frac{d^{2}p}{(2\pi)^{2}}\frac{d^{2}q}{(2\pi)^{2}}
  \Big<\widetilde{\psi}_{\beta}(p)
  \widetilde{\bar{\psi}}_{\alpha}(q)
 \Big>e^{ip.(x+\frac{L}{2}-\frac{\epsilon}{2})}e^{-iq.(x+\frac{L}{2}+\frac{\epsilon}{2})}}
 \nonumber\\
 &=&(\gamma^{\mu}\gamma^{5})^{\alpha\beta}\int{\frac{d^{2}p}{(2\pi)
 ^{2}}\Big(\frac{i\PS}{p^{2}}\Big)_{\beta\alpha}e^{-ip.\epsilon}}\nonumber\\
 &=&-\frac{i}{2\pi}\mbox{Tr}(\gamma^{\rho}\gamma^{\mu}\gamma^{5})\frac{\epsilon_{\rho}}
 {\epsilon^{2}}.
 \end{eqnarray}
Using as next
 $\lim\limits_{\epsilon\longrightarrow0}\frac{\epsilon_{\rho}\epsilon^{\nu}}{\epsilon^{2}}
 =\frac{1}{d}\delta^{\nu}_{\rho}$, where $d$ is the space-time
 dimension, and
 $\mbox{Tr}(\gamma^{\rho}\gamma^{\mu}\gamma^{5})=2\varepsilon^{\rho\mu}$,
and replacing the result from Eq. (\ref{S39}) in the expression
on the r.h.s. of Eq. (\ref{S38}), we arrive at the axial anomaly
of two dimensional noncommutative dipole QED with adjoint
matters, which reads:
\begin{eqnarray}\label{S40}
 \Big<\partial^{\mu}J_{\mu(5)}(x)\Big>=\frac{g}{2\pi}\varepsilon^{\mu\nu}
 \Big[F_{\mu\nu}(x+L)+F_{\mu\nu}(x-L)-2F_{\mu\nu}(x)\Big].
\end{eqnarray}
Since the field strength tensor is dipoleless, the above result is
gauge invariant. Further the anomaly vanishes, if we integrate
both sides over one space coordinate $x^{1}$:
\begin{eqnarray}\label{S41}
 \int{dx^{1}\Big<\partial^{\mu}J_{\mu(5)}(x)\Big>}=0.
\end{eqnarray}
Note that classically the axial charge $Q_{5}$,
\begin{eqnarray}\label{S42}
Q_{5}\equiv \int dx^{1} J_{0(5)}(x),
\end{eqnarray}
corresponding to $J_{\mu(5)}$ from Eq. (\ref{S32}) vanishes. The
above result shows that in the quantum level $Q_{5}$ is still
conserved.
\newpage
\par\noindent {\it ii) Four Dimensions} \vskip0.3cm\par\noindent
To obtain the axial anomaly in four dimensions using the point
split regularization method, the same steps leading from Eq.
(\ref{S32}) to Eq. (\ref{S38}) must be repeated. It turns out that
in the zeroth order of the perturbative expansion the contribution
to the VEV of two matter fields from Eq. (\ref{S38}) vanishes. In
the first order of perturbative expansion we have therefore
\begin{eqnarray}\label{S45}
\lefteqn{\Big<\partial^{\mu}J_{\mu(5)}(x)\Big>=}\nonumber\\
 &&=
 \mbox{symm}\lim_{\epsilon\longrightarrow0}g\epsilon^{\nu}
 \Biggr\{2\Big[\tau^{\mu}_{1}(x,\epsilon)-\tau^{\prime\mu}_{1}
 (x,\epsilon)\Big]\partial_{\nu}A_{\mu}(x)
 +\tau^{\mu}_{1}(x,\epsilon)F_{\mu\nu}(x+L)-\tau^{\prime\mu}_{1}
 (x,\epsilon)F_{\mu\nu}(x-L)\Biggr\},\nonumber\\
\end{eqnarray}
where two functions $\tau_{1}^{\mu}(x,\epsilon)$ and
$\tau_{1}^{'\mu}(x,\epsilon)$ are defined by:
\begin{eqnarray}\label{S43}
 \tau^{\mu}_{1}(x,\epsilon)\equiv(\gamma^{\mu}\gamma^{5})^{\alpha\beta}
 \Big<\psi_{\beta}(x-\frac{\epsilon}{2})\star
 \bar{\psi}_{\alpha}(x+\frac{\epsilon}{2})(ig)\int{d^{4}z\Big(\bar{\psi}\gamma^{\lambda}\star
 [A_{\lambda},\psi]_{\star}\Big)(z)}\Big>,
\end{eqnarray}
and
\begin{eqnarray}\label{S44}
 \tau'^{\mu}_{1}(x,\epsilon)\equiv(\gamma^{\mu}\gamma^{5})^{\alpha\beta}
 \Big<\bar{\psi}_{\alpha}(x+\frac{\epsilon}{2})
 \star\psi_{\beta}(x-\frac{\epsilon}{2})(ig)\int{d^{4}z\Big(\bar{\psi}\gamma^{\lambda}\star
 [A_{\lambda},\psi]_{\star}\Big)(z)}\Big>.
\end{eqnarray}
The function $\tau_{1}(x,\epsilon)$ can be evaluated using
standard perturbative methods and is given by:
\begin{eqnarray}\label{S46}
\lefteqn{\hspace{-0.5cm}
 \tau^{\mu}_{1}(x,\epsilon)=-ig\int{\frac{d^{4}p}{(2\pi)^{4}}\frac{d^{4}q}{(2\pi)^{4}}
 \frac{d^{4}k_{1}}{(2\pi)^{4}}
 \frac{d^{4}k_{2}}{(2\pi)^{4}}\frac{d^{4}k}{(2\pi)^{4}}\int{d^{4}ze^{ip.(x+\frac{L}{2}-\frac{\epsilon}{2})}
 e^{-iq.(x+\frac{L}{2}+\frac{\epsilon}{2})}e^{ik.z}}}
 }\nonumber\\
&&\hspace{2cm}\times
(\gamma^{\lambda})^{\rho\sigma}(\gamma^{\mu}\gamma^{5})^{\alpha\beta}\widetilde{A}_{\lambda}(k)\Biggr[\Big<
 \widetilde{\psi}_{\beta}(p)\widetilde{\bar{\psi}}_{\alpha}(q)\widetilde{\psi}_{\sigma}(k_{2})\widetilde{\bar{\psi}
 }_{\rho}(k_{1})\Big>e^{ik_{2}.(z+\frac{L}{2})}e^{-ik_{1}.(z+\frac{L}{2})}\nonumber\\
&& \hspace{6cm}
+\Big<\widetilde{\psi}_{\beta}(p)\widetilde{\bar{\psi}}_{\alpha}(q)\widetilde{\bar{\psi}}_{\rho}(k_{1})\widetilde{\psi}_
 {\sigma}(k_{2})\Big>e^{-ik_{1}.(z-\frac{L}{2})}e^{ik_{2}.(z-\frac{L}{2})}\Biggr]\nonumber\\
&&=-ig(\gamma^{\lambda})^{\rho\sigma}(\gamma^{\mu}\gamma^{5})^{\alpha\beta}
 \int{\frac{d^{4}k}{(2\pi)^{4}}\frac{d^{4}k_{2}}{(2\pi)^{4}}\frac{(\KS+\KS_{2})_{\beta\rho}}{(k+k_{2})^{2}}
 \frac{(\KS_{2})_{\sigma\alpha}}{k_{2}^{2}}\widetilde{A}_{\lambda}(k)e^{ik.(x-\frac{\epsilon}{2})}e^{-ik_{2}
 .\epsilon}\big(1-e^{ik.L}\big)}.
\end{eqnarray}
For the limit $\epsilon\to 0$ and  for large $k_{2}$ the last
integral reads
\begin{eqnarray}\label{S47}
 \tau^{\mu}_{1}(x,\epsilon)&=&-ig(\gamma^{\lambda})^{\rho\sigma}
 (\gamma^{\mu}\gamma^{5})^{\alpha\beta}\Big(\int{\frac
 {d^{4}k_{2}}{(2\pi)^{4}}\frac{k_{2\zeta}}{k_{2}^{4}}
 (\gamma^{\zeta})_{\sigma\alpha}e^{-ik_{2}.\epsilon}}\Big)
 \nonumber\\
 &&\times\Biggr[\int{\frac{d^{4}k}{(2\pi)^{4}}k_{\delta}
 (\gamma^{\delta})_{\beta\rho}\widetilde{A}_{\lambda}(k)
 e^{ik.(x-\frac{\epsilon}{2})}}-\int{\frac{d^{4}k}{(2\pi)^{4}}k_{\delta}
 (\gamma^{\delta})_{\beta\rho}
 \widetilde{A}_{\lambda}(k)e^{ik.(x+L-\frac{\epsilon}{2})}}\Biggr]\nonumber\\
 &&=\frac{g}{8\pi^{2}}\frac{\epsilon_{\zeta}}{\epsilon^{2}}\Big(\partial_{\delta}A_{\lambda}
 (x-\frac{\epsilon}{2})-\partial_{\delta}A_{\lambda}(x+L-\frac{\epsilon}{2})\Big)\mbox{Tr}(\gamma^{5}\gamma^{\mu}
 \gamma^{\delta}\gamma^{\lambda}\gamma^{\zeta}).
\end{eqnarray}
Similarly $\tau'^{\mu}_{1}(x,\epsilon)$ can be easily evaluated
and is given by:
\begin{eqnarray}\label{S48}
  \tau^{\prime\mu}_{1}(x,\epsilon)=-\frac{ig}{2\pi^{2}}
  \varepsilon^{\mu\delta\lambda\zeta}\frac{\epsilon_{\zeta}}
  {\epsilon^{2}}\Big(\partial_{\delta}A_{\lambda}
 (x-\frac{\epsilon}{2})-\partial_{\delta}A_{\lambda}(x-L-\frac{\epsilon}{2})\Big),
\end{eqnarray}
where
$\mbox{Tr}(\gamma^{5}\gamma^{\mu}\gamma^{\delta}\gamma^{\lambda}\gamma^{\zeta})
=-4i\varepsilon^{\mu\delta\lambda\zeta}$ is used. Replacing now
the results from Eqs. (\ref{S47}) and (\ref{S48}) in Eq.
(\ref{S45}), we arrive at the axial anomaly of the four
dimensional noncommutative dipole QED with adjoint matters:
\begin{eqnarray}\label{S49}
 \Big<\partial^{\mu}J_{\mu(5)}(x)\Big>&=&
\frac{g^{2}}{16\pi^{2}}\varepsilon^{\mu\nu\lambda\delta}\Biggr[F
 _{\mu\nu}(x-L)F_{\delta\lambda}(x-L)-F_{\mu\nu}(x+L)F_{\delta\lambda}(x+L)\nonumber\\
&&
-2\Big(F_{\mu\nu}(x+L)-F_{\mu\nu}(x-L)\Big)F_{\delta\lambda}(x)\Biggr].
\end{eqnarray}
The same result is also obtained in \cite{h12} where the
Fujikawa's path integral  method  is used. As in two dimensional
case, the above result turns out to be  gauge invariant and
vanishes after integrating over three spacial coordinates:
\begin{eqnarray}\label{S50}
 \int{d^{3}x\Big<\partial^{\mu}J_{\mu(5)}(x)\Big>}=0.
\end{eqnarray}
\subsubsection{Triangle Anomaly} In this section the triangle
diagrams will be calculated in two and four dimensions using the
dimensional regularization method. \vskip0.3cm\par\noindent {\it
i) Two  Dimensions} \vskip0.3cm\par\noindent Let us consider the
two point function:
\begin{eqnarray}\label{S51}
 \Gamma^{\mu\nu}(x,y)\equiv\Big<T\Big(J^{\mu(5)}(x) J^{\nu}(y)\Big)\Big>,
\end{eqnarray}
with $J^{\nu}(x)$ and $ J^{\mu(5)}(y)$ defined in Eqs. (\ref{S30})
and (\ref{S32}),  respectively. After an appropriate shift of
integration variable the dimensional regulated Feynman integral
corresponding to the above two point function is given by:
\begin{eqnarray}\label{S52}
\Gamma^{\mu\nu}(x,y)=g\int{\frac{d^{d}q}{(2\pi)^{d}}e^{iq.(x-y)}
\big(2-e^{iq.L}-e^{-iq.L}\big)}\int{\frac{d^{d}\ell}
 {(2\pi)^{d}}\mbox{Tr}\Big(\frac{1}{\LS}\gamma^{\mu}\gamma^{5}
 \frac{1}{\LS+\QS}\gamma^{\nu}\Big)}.
\end{eqnarray}
The divergence of $\Gamma^{\mu\nu}$ with respect to $y^{\nu}$ can
be easily calculated. As it turns out the vector Ward identity
vanishes:
\begin{eqnarray}\label{S53}
\Big<\frac{\partial}{\partial y^{\nu}}J^{\nu}(y)\Big>=0.
\end{eqnarray}
What concerns the axial vector Ward identity, the divergence of
$\Gamma^{\mu\nu}(x,y)$ with respect to $x$ is to be calculated.
It is given by:
\begin{eqnarray}\label{S54}
 \frac{\partial}{\partial x^{\mu}}\Gamma^{\mu\nu}(x,y)\propto
 \int{\frac{d^{d}\ell}{(2\pi)^{d}}\mbox{Tr}
 \Big(\frac{1}{\LS}\QS\gamma^{5}\frac{1}{\LS+\QS}\gamma^{\nu}\Big)}.
\end{eqnarray}
Using the standard 't Hooft's definition of $\gamma_{5}$ in
d-dimensions and the following identity:
\begin{eqnarray}\label{S55}
 \QS\gamma^{5}=-\gamma^{5}(\QS+\LS)-\LS\gamma^{5}+2\gamma^{5}\LS_{\bot},
\end{eqnarray}
we arrive at:
\begin{eqnarray}\label{S56}
 \frac{\partial}{\partial x^{\mu}}\Gamma^{\mu\nu}(x,y)=-\frac{g}{\pi}\varepsilon^{\alpha\nu}
 \int{\frac{d^{2}q}{(2\pi)^{2}}e^{iq.(x-y)}q_{\alpha}\big(2-2\cos(q.L)\big)}.
\end{eqnarray}
Now using the relation
\begin{eqnarray}\label{S57}
 \Big<\partial^{\mu}J_{\mu(5)}(x)\Big>=\int{d^{2}y\ \frac{\partial}{\partial
 x^{\mu}}\Gamma^{\mu\nu}(x,y)
 A_{\nu}(y)},
\end{eqnarray}
and the result from Eq. (\ref{S56}) we obtain:
\begin{eqnarray}\label{S58}
 \Big<\partial^{\mu}J_{\mu(5)}(x)\Big>=
 \frac{g}{2\pi}\varepsilon^{\alpha\beta}\Big(F_{\alpha\beta}(x+L)+F_{\alpha\beta}(x-L)
 -2F_{\alpha\beta}(x)\Big),
\end{eqnarray}
which coincides with our result from the point split
regularization [see Eq. (\ref{S40})]. \vskip0.3cm\par\noindent
{\it ii) Four Dimensions}
\par\noindent\vskip0.3cm
To calculate the anomaly in four dimensions, the three point
function of one axial vector and two vector currents must be
considered:
\begin{eqnarray}\label{S59}
 \Gamma_{\mu\lambda\nu}(x,y,z)\equiv\bigg<T\Big(J_{\mu(5)}(x)
J_{\lambda}(y)J_{\nu}(z)\Big)\bigg>.
\end{eqnarray}
The triangle Feynman integrals corresponding to the
$\Gamma_{\mu\lambda\nu}$ is then given by:
\begin{eqnarray}\label{S60}
\lefteqn{
\Gamma_{\mu\lambda\nu}(x,y,z)=}\nonumber\\
&&=
2g^{2}\int{\frac{d^{d}k_{2}}{(2\pi)^{d}}\frac{d^{d}k_{3}}{(2\pi)^{d}}e^{-i(k_{2}+k_{3}).x}
 e^{ik_{2}.y}e^{ik_{3}.z}}\Big[\sin\big(k_{2}.L\big)+\sin\big(k_{3}.L\big)-
 \sin\big((k_{2}+k_{3}).L\big)\Big]\nonumber\\
&&\ \ \times
\int{\frac{d^{d}\ell}{(2\pi)^{d}}\Big[\mbox{Tr}\Big(\frac{1}{\LS+\KS_{3}}
\gamma^{\mu}\gamma^{5}\frac{1}{\LS-\KS_{2}}
 \gamma^{\lambda}\frac{1}{\LS}\gamma^{\nu}\Big)+\Big((k_{2},\lambda)
 \leftrightarrow(k_{3},\nu)\Big)\Big]}.
\end{eqnarray}
The divergence of the above integral with respect to
$y^{\lambda}$ can be easily calculated, leading to vanishing
vector Ward identity:
\begin{eqnarray*}
 \Big<\frac{\partial}{\partial y^{\lambda}}J^{\lambda}(y)\Big>=0.
\end{eqnarray*}
Further, the divergence of $\Gamma_{\mu\lambda\nu}$ with respect
to $x^{\mu}$ is given by:
\begin{eqnarray}\label{S61}
\lefteqn{\frac{\partial}{\partial
x^{\mu}}\Gamma^{\mu\lambda\nu}(x,y,z)=}\nonumber\\
&&2ig^{2}\int{\frac{d^{d}k_{2}}{(2\pi)^{d}}\frac{d^{d}k_{3}}{(2\pi)^{d}}e^{-i(k_{2}+k_{3}).x}
 e^{ik_{2}.y}e^{ik_{3}.z}}\Big[\sin\big(k_{2}.L\big)+\sin\big(k_{3}.L\big)-
 \sin\big((k_{2}+k_{3}).L\big)\Big]\nonumber\\
&&\times
\bigg[R^{\lambda\nu}(k_{2},k_{3})+A^{\lambda\nu}(k_{2},k_{3})\bigg],
\end{eqnarray}
where two functions $R^{\lambda\nu}$ and $A^{\lambda\nu}$ are
defined by:
\begin{eqnarray}\label{S62}
 R^{\lambda\nu}(k_{2},k_{3})&=&\int{\frac{d^{d}\ell}{(2\pi)^{d}}
 \Biggr[\mbox{Tr}\Big(\gamma^{5}\frac{1}{\LS-\KS_{2}}
 \gamma^{\lambda}\frac{1}{\LS}\gamma^{\nu}\Big)+
 \mbox{Tr}\Big(\frac{1}{\LS+\KS_{3}}\gamma^{5}\gamma^{\lambda}\frac{1}{\LS}\gamma^{\nu}\Big)}
 \nonumber\\
 &&+\mbox{Tr}\Big(\gamma^{5}\frac{1}{\LS-\KS_{3}}\gamma^{\nu}\frac{1}{\LS}\gamma^{\lambda}\Big)+
 \mbox{Tr}\Big(\frac{1}{\LS+\KS_{2}}\gamma^{5}\gamma^{\nu}\frac{1}{\LS}\gamma^{\lambda}\Big)
 \Biggr],
 \end{eqnarray}
and
\begin{eqnarray}\label{S63}
 A^{\lambda\nu}(k_{2},k_{3})=\int{\frac{d^{d}\ell}{(2\pi)^{d}}\Biggr[\mbox{Tr}
 \Big(\frac{1}{\LS+\KS_{3}}\gamma^{5}\LS_{\bot}\frac{1}{\LS-\KS_{2}}\gamma^{\lambda}
 \frac{1}{\LS}\gamma^{\nu}\Big)+\mbox{Tr}\Big((k_{2},\lambda)\leftrightarrow
 (k_{3},\nu)\Big)\Biggr]}.
\end{eqnarray}
Here, we have used the identity $
 (\KS_{2}+\KS_{3})\gamma^{5}=(\LS+\KS_{2})\gamma^{5}+
 \gamma^{5}(\LS-\KS_{3})-2\gamma^{5}\LS_{\bot}$. As it turns out the function $R^{\lambda\nu}$
vanishes using the cyclic permutation symmetry of the trace. The
anomaly is then entirely given by  $A^{\lambda\nu}$, which reads
\begin{eqnarray}\label{S64}
 A^{\lambda\nu}(k_{2},k_{3})=
 \frac{1}{4\pi^{2}}\varepsilon^{\alpha\beta\lambda\nu}k_{2\alpha}k_{3\beta}.
\end{eqnarray}
Replacing this results in Eq. (\ref{S61}) and using the identity
\begin{eqnarray}\label{S65}
 \Big<\partial^{\mu}J_{\mu(5)}(x)\Big>=
 \frac{1}{2}\int{d^{4}yd^{4}z\frac{\partial}{\partial x^{\mu}}\Gamma^{\mu\lambda\nu}(x,y,z)
 A_{\lambda}(y)A_{\nu}(z)},
\end{eqnarray}
the VEV of the divergence of $J_{\mu(5)}$ is given by the same Eq.
(\ref{S49}), which we have obtained by making use of point
splitting regularization method.
\subsection{The Axial Anomaly of $J'_{\mu(5)}$ and
$J''_{\mu(5)}$} As we have seen above, the noncommutative QED with
adjoint matters consists three different vector and axial vector
currents. In this section we will calculate the axial anomaly
corresponding to two currents $J'_{\mu(5)}$ and $J''_{\mu(5)}$
from Eq. (\ref{S33}) in two and four dimensions by making use of
dimensional regularization. \vskip0.3cm
\par\noindent{\it i)} Two Dimensions \vskip0.3cm\par\noindent
Let us indicate the two-point function corresponding to these
currents by $\Gamma'_{\mu\nu}$ and $\Gamma''_{\mu\nu}$
respectively:
\begin{eqnarray}\label{S66}
 \Gamma'_{\mu\nu}(x,y)=\Big<T\Big(J'_{\mu(5)}(x)J_{\nu}(y)\Big)\Big>,\hspace{1cm}
 \Gamma''_{\mu\nu}(x,y)=\Big<T\Big(J''_{\mu(5)}(x)J_{\nu}(y)\Big)\Big>.
\end{eqnarray}
In a dimensional regularization, the Feynman integrals
corresponding to these two-point functions are given by:
\begin{eqnarray}\label{S67}
 \Gamma'^{\mu\nu}(x,y)=g\int{\frac{d^{d}q}{(2\pi)^{d}}e^{iq.(x-y)}\big(1-e^{iq.L}\big)}\int{
 \frac{d^{d}\ell}{(2\pi)^{d}}\mbox{Tr}\Big(\frac{1}{\LS}\gamma^{\mu}\gamma^{5}
 \frac{1}{\LS+\QS}\gamma^{\nu}\Big)},
\end{eqnarray}
and
\begin{eqnarray}\label{S68}
 \Gamma''^{\mu\nu}(x,y)=g\int{\frac{d^{d}q}{(2\pi)^{d}}e^{iq.(x-y)}\big(1-e^{-iq.L}\big)}\int{
 \frac{d^{d}\ell}{(2\pi)^{d}}\mbox{Tr}\Big(\frac{1}{\LS}\gamma^{\mu}\gamma^{5}\frac{1}{\LS+\QS}
 \gamma^{\nu}\Big)}.
\end{eqnarray}
Deriving the above integrals with respect to $x^{\mu}$ and going
through the same standard procedure leading to the divergence of
the axial vector currents, we arrive at:
\begin{eqnarray}\label{S69}
 \Big<\partial^{\mu}J'_{\mu(5)}\Big>=\frac{g}{2\pi}\varepsilon^{\alpha\beta}
 \Big(F_{\alpha\beta}(x+L)-F_{\alpha\beta}
 (x)\Big),\hspace{1cm}
  \Big<\partial^{\mu}J''_{\mu(5)}\Big>=\frac{g}{2\pi}\varepsilon^{\alpha\beta}
 \Big(F_{\alpha\beta}(x-L)-F_{\alpha\beta}(x)\Big).
\end{eqnarray}
Integrating now these results over $x^{1}$ we obtain:
\begin{eqnarray}\label{S70}
 \int{dx^{1}\Big<\partial^{\mu}J'_{\mu(5)}\Big>}=0,\hspace{1cm}\mbox{and}\hspace{1cm}
  \int{dx^{1}\Big<\partial^{\mu}J''_{\mu(5)}\Big>}=0.
\end{eqnarray}
The corresponding axial charges are therefore conserved.
\vskip0.3cm\par\noindent {\it ii) Four Dimensions}
\vskip0.3cm\par\noindent Let us now consider the three-point
function corresponding to the axial vector currents $J'_{\mu(5)}$
and $J''_{\mu(5)}$ from Eq. (\ref{S33}):
\begin{eqnarray}\label{S71}
 \Gamma'_{\mu\lambda\nu}(x,y,z)=\Big<T\Big(J'_{\mu(5)}(x)J_{\lambda}(y)J_{\nu}(z)\Big)\Big>,
 \hspace{0.5cm}
 \Gamma''_{\mu\lambda\nu}(x,y,z)=\Big<T\Big(J''_{\mu(5)}(x)J_{\lambda}(y)J_{\nu}(z)\Big)\Big>.
\end{eqnarray}
The dimensional regularized Feynman integrals are given by:
\begin{eqnarray*}
 \Gamma'^{\mu\lambda\nu}(x,y,z)&=&ig^{2}\int{\frac{d^{d}k_{2}}{(2\pi)^{d}}
 \frac{d^{d}k_{3}}{(2\pi)^{d}}e^{-i(k_{2}+k_{3}).x}
 e^{ik_{2}.y}e^{ik_{3}.z}}\big(1-e^{-ik_{2}.L}\big)\big(1-e^{-ik_{3}.L}\big)\nonumber\\
&&\times\int\frac{d^{d}\ell}{(2\pi)^{d}}\Big[\mbox{Tr}\Big(\frac{1}{\LS+\KS_{3}}
\gamma^{\mu}\gamma^{5}\frac{1}{\LS-\KS_{2}}
 \gamma^{\lambda}\frac{1}{\LS}\gamma^{\nu}\Big)+\Big((k_{2},\lambda)
 \leftrightarrow(k_{3},\nu)\Big)\Big],
\end{eqnarray*}
and
\begin{eqnarray*}
 \Gamma''^{\mu\lambda\nu}(x,y,z)&=&ig^{2}\int{\frac{d^{d}k_{2}}{(2\pi)^{d}}
 \frac{d^{d}k_{3}}{(2\pi)^{d}}e^{-i(k_{2}+k_{3}).x}
 e^{ik_{2}.y}e^{ik_{3}.z}}\big(1-e^{+ik_{2}.L}\big)\big(1-e^{+ik_{3}.L}\big)\nonumber\\
&&\times
\int\frac{d^{d}\ell}{(2\pi)^{d}}\Big[\mbox{Tr}\Big(\frac{1}{\LS+\KS_{3}}
\gamma^{\mu}\gamma^{5}\frac{1}{\LS-\KS_{2}}
\gamma^{\lambda}\frac{1}{\LS}\gamma^{\nu}\Big)+\Big((k_{2},\lambda)
\leftrightarrow(k_{3},\nu)\Big)\Big].
\end{eqnarray*}
The axial vector anomalies corresponding to these currents can be
calculated using the standard dimensional regularization
procedure and read
\begin{eqnarray}\label{S72}
\Big<\partial^{\mu}J'_{\mu(5)}\Big>=\frac{g^{2}}{16\pi^{2}}
 \varepsilon^{\alpha\beta\lambda\nu}\Big[F_{\alpha\lambda}(x)
 F_{\beta\nu}(x)+F_{\alpha\lambda}(x+L)F_{\beta\nu}(x+L)
 -2F_{\alpha\lambda}(x)F_{\beta\nu}(x+L)\Big],
 \end{eqnarray}
 and
 \begin{eqnarray}\label{S73}
 \Big<\partial^{\mu}J''_{\mu(5)}\Big>=-\frac{g^{2}}{16\pi^{2}}
 \varepsilon^{\alpha\beta\lambda\nu}\Big[F_{\alpha\lambda}(x)
 F_{\beta\nu}(x)+F_{\alpha\lambda}(x-L)F_{\beta\nu}(x-L)
 -2F_{\alpha\lambda}(x)F_{\beta\nu}(x-L)\Big],
 \end{eqnarray}
 respectively. Integrating over three spacial coordinates and using an
appropriate shift of integration variable, we arrive at:
\begin{eqnarray}\label{S74}
 \int{d^{3}x\Big<\partial^{\mu}J'_{\mu(5)}\Big>}&=& -\int{d^{3}x\Big<\partial^{\mu}
 J''_{\mu(5)}\Big>}\nonumber\\
 &=&\frac{g^{2}}{8\pi^{2}}\varepsilon^{\alpha\beta\lambda\nu}\Big(
 \int{d^{3}xF_{\alpha\lambda}(x)F_{\beta\nu}(x)}-\int{d^{3}xF_{\alpha\lambda}(x-\frac{L}{2})
 F_{\beta\nu}(x+\frac{L}{2})}\Big),
\end{eqnarray}
which, in contrary to the result for $J_{\mu(5)}$ from Eq.
(\ref{S50}) does not vanish. This means that the corresponding
axial charges to $J'_{\mu(5)}$ and $J''_{\mu(5)}$ are anomalous.
\section{One-Loop Perturbative Calculation} \setcounter{equation}{0} In this section, a
perturbative calculation is performed to obtain the one-loop
contribution to the renormalization constants $Z_{i}$, $i=1,2,3$
of a dipole U(1) gauge theory with matter fields in the adjoint
representation. \vspace{0.5cm}\par\noindent{\it i) Fermion Self
Energy}\vspace{0.5cm}\par\noindent Using the vertex of two Fermion
and one gauge  field of the noncommutative dipole U(1) gauge
theory from Eq. (\ref{S21}), the Feynman integral corresponding to
the one-loop fermion self energy diagram \vspace{1cm}
\begin{figure}[ht]
\begin{center}
\SetScale{1}
    \begin{picture}(50,20)(0,0)
    \Text(-80,0)[]{$-i\Sigma(\PS)\ \ \equiv$}
    \Vertex(0,0){1.5}
    \Vertex(60,0){1.5}
    \ArrowLine(-40,0)(0,0)
    \ArrowLine(0,0)(60,0)
    \ArrowLine(60,0)(100,0)
    \Text(-20,-10)[]{$p$}
    \Text(30,-10)[]{$p-k$}
    \Text(30,40)[]{$k$}
    \Text(80,-10)[]{$p$}
    \PhotonArc(30,0)(30,0,180){3}{8}
    \end{picture}
\end{center}
\caption{The diagram contributing to the one-loop fermion self
energy.}
\end{figure}
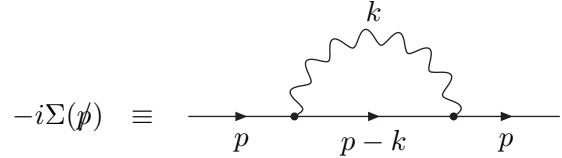
 can be given by
\begin{eqnarray}\label{S75}
\Sigma(\PS)=-4ig^{2}\int{\frac{d^{d}k}{(2\pi)^{d}}
\gamma_{\mu}\frac{\PS-\KS+m}{k^{2}\big[(p-k)^{2}-m^{2}\big]}
 \gamma^{\mu}\sin^{2}\big(\frac{k.L}{2}\big)}.
\end{eqnarray}
Using now the identity $\sin^{2}x= \frac{1}{2}(1-\cos (2x))$, it
can be shown that the above integral includes planar and nonplanar
parts. The planar part is exactly twice the value of the one-loop
contribution of fermion self energy in the {\it commutative} U(1)
gauge theory. In the minimal subtraction (MS) scheme, the mass and
wave function renormalization constants $\delta m$ and $Z_{2}$ are
therefore given by:
\begin{eqnarray}\label{S76}
\delta{m}=\Sigma(\PS)\big|_{\PS=m}=-\frac{3mg^{2}}{4\pi^{2}}\
\frac{1}{(4-d)} ,\hspace{1cm}Z_{2}\equiv 1+\delta
Z_{2}\hspace{1cm}\mbox{with}\hspace{1cm}\delta
Z_{2}=\frac{g^{2}}{4\pi^{2}}\ \frac{1}{(4-d)}.
\end{eqnarray}
What concerns the nonplanar part, it is considered to be finite
for finite dipole length $L$. In the $L\to 0$ limit, a UV/IR
mixing will occur. In this case, the infrared divergences which
arise from the nonplanar part cancel the ultraviolet divergences
from the planar part, leading to a vanishing renormalization
constant and eventually to a vanishing $\beta$-function. This is
indeed consistent with the observation that the theory is free in
the limit $L\to 0$.
 \vspace{0.5cm}\par\noindent{\it ii) Vertex
Function}\vspace{0.5cm}\par\noindent The one-loop vertex function
in noncommutative dipole QED, receives contribution from only one
vertex diagram: \vspace{1cm}
\begin{figure}[ht]
\begin{center}
\SetScale{1}
   \begin{picture}(30,10)(0,0)
    \Vertex(100,0){1.5}
    \Photon(100,0)(100,40){2}{4}
    \Text(100,-35)[]{$k$}
    \Photon(75,-25)(125,-25){2}{4}
    \LongArrow(105,20)(105,10)
    \ArrowLine(100,0)(75,-25)
    \ArrowLine(125,-25)(100,0)
    \Vertex(125,-25){1.5}
    \Vertex(75,-25){1.5}
    \ArrowLine(75,-25)(50,-50)
    \ArrowLine(150,-50)(125,-25)
    \Text(120,15)[]{$q,\mu$}
    \Text(50,-40)[]{$p'$}
    \Text(150,-40)[]{$p$}
    \Text(135,-25)[]{$\rho$}
    \Text(65,-25)[]{$\sigma$}
    \Text(-20,0)[]{$\bar{u}(p^{\prime})\Big(-2g\
    \delta\Gamma^{\mu}(p^{\prime},p)\
    \sin\big(\frac{q.L}{2}\big)\Big)u(p)\ \ \equiv$}
    \end{picture}
\end{center}
\vskip1cm \caption{The diagram contributing to the one-loop vertex
function.}
\end{figure}
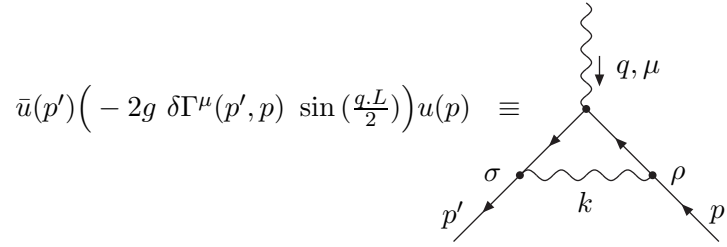
\vskip0.5cm
\par\noindent where $\delta\Gamma^{\mu}$ is the one-loop
contribution to the vertex function.  Remember that in the Moyal
case, due to the appearance of additional three gauge vertex, a
second diagram contributes to the one-loop vertex function of the
theory. Using the Feynman rules of noncommutative dipole U(1)
gauge theory with adjoint matters, the Feynman integral
corresponding to the above diagram can be given by:
\begin{eqnarray}\label{S77}
 \delta\Gamma^{\mu}(p^{\prime},p)=-4ig^{2}\int{\frac{d^{d}k}{(2\pi)^{d}}
 \frac{\gamma_{\rho}
 (\PS^{\prime}-\KS+m)\gamma^{\mu}(\PS-\KS+m)\gamma^{\rho}}{k^{2}
 \big[(p^{\prime}-k)^{2}-m^{2}\big]
 \big[(p-k)^{2}-m^{2}\big]}\sin^{2}\big(\frac{k.L}{2}\big)}.
\end{eqnarray}
Replacing the factor $\sin^{2}(\frac{k\cdot L}{2})$ by
$\frac{1}{2}(1-\cos (k\cdot L))$ in the above integral the planar
and the nonplanar parts of the above integral can be separated.
Its planar part is twice the vertex integral in the commutative
case and its nonplanar part turns out to be finite for finite
value of $L$. If we use the standard mimimal subtraction (MS)
scheme, where only UV divergences from the planar diagrams are to
be taken into account, the wave function renormalization constant
can be obtained and reads:
\begin{eqnarray}\label{S78}
Z_{1}\equiv 1+\delta
Z_{1}\hspace{1cm}\mbox{with}\hspace{1cm}\delta Z_{1}=
\frac{g^{2}}{4\pi^{2}}\ \frac{1}{(4-d)}.
\end{eqnarray}
Comparing this result with the result of $Z_{2}$ from Eq.
(\ref{S76}), it turns out that the identity $Z_{1}=Z_{2}$ is
satisfied in the MS scheme.
\par
Now let us consider again the Feynman integral corresponding to
the fermion self-energy from Eq. (\ref{S75}). Using the usual
definition of $\delta Z_{2} \equiv
\frac{d\Sigma(\PS)}{d\PS}\big|_{\PS=m}$ we have
\begin{eqnarray}\label{S79}
 \delta Z_{2}=
 8ig^{2}\Big[\int{\frac{d^{4}k}{(2\pi)^{4}}\frac{(k-p)^{2}-4p.(k-p)+m^{2}}
 {k^{2}\big[(k-p)^{2}-m^{2}\big]^{2}}\sin^{2}\big(\frac{k.L}{2}\big)}\Big]_{p=(m,0)},
\end{eqnarray}
which turns out to be equal to $\delta Z_{1}$ coming from the
vertex function. The identity $Z_{1}=Z_{2}$ seems therefore to be
valid for both planar and nonplanar parts. This is one of the
results from the Ward-Takahashi (WT) identity, which will be
proved in the next section. We will show its validity in any order
of perturbative expansion and any renormalization prescription
scheme.
\par\vspace{0.5cm}\par\noindent{\it iii) Photon Self-Energy}
\vspace{0.5cm}\par\noindent\vspace{0.5cm} The Feynman integral
corresponding to the photon self-energy diagram \vskip0.2cm
\begin{figure}[ht]
\begin{center}
\SetScale{1}
    \begin{picture}(50,20)(0,0)
    \Text(-80,0)[]{$i\Pi^{\mu\nu}(p)\ \ \equiv$}
    \Vertex(0,0){1.5}
    \Vertex(60,0){1.5}
    \Photon(-40,0)(0,0){3}{4}
    \Photon(60,0)(100,0){3}{4}
    \Text(-20,-10)[]{$p$}
    \Text(30,-40)[]{$k$}
    \Text(80,-10)[]{$p$}
    \LongArrowArc(30,0)(30,0,360)
    \end{picture}
\end{center}
\vskip0.5cm \caption{The diagram contributing to the one-loop
photon self-energy}
\end{figure}
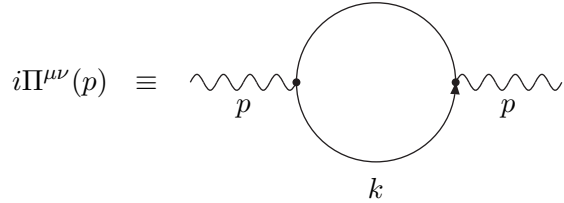
\vskip0.5cm
\par\noindent
is given by:
\begin{eqnarray}\label{S79x}
i\Pi^{\mu\nu}_{2}(p)=+4g^{2}\sin^{2}\big(\frac{p.L}{2}\big)\int\frac{d^{d}k}{(2\pi)^{d}}\mbox{Tr}
\Biggr[\gamma^{\mu}\frac{i}{\KS-m}\gamma^{\nu}
 \frac{i}{\KS-\PS-m}\Biggr].
\end{eqnarray}
This integral involves only a planar part. The planar phase factor
appears only as a multiplicative factor before the Feynman
integral which is exactly the same integral from the commutative
U(1) and turns out to be proportional to usual tensorial structure
$(\eta_{\mu\nu}p^{2}-p_{\mu}p_{\nu})$. As in the commutative U(1)
the coupling constant renormalization parameter is defined for
small external momenta $p$. In this limit, it is given by:
\begin{eqnarray}\label{S80}
Z_{3}=1+\delta Z_{3},\hspace{1cm}\mbox{with}\hspace{1cm}\delta
Z_{3}= \frac{g^{2}}{6\pi^{2}}\ \frac{1}{(4-d)}(p\cdot L)^{2}.
\end{eqnarray}
In the next section, after proving the Ward-Takahashi identity for
noncommutative dipole QED with adjoint matters, the general form
of $Z_{3}$ will be presented. We will show that the above result
is valid for all orders of perturbative expansion.
\section{Ward-Takahashi Identity of Noncommutative Dipole QED} \setcounter{equation}{0}
In this section the Ward-Takahashi identity of the noncommutative
dipole QED associated to the gauge invariance of the theory is
proved for all order of perturbative expansion. The matter fields
are still in the adjoint representations. As in the commutative
QED, there are two different methods to show this identity: the
perturbative and the path integral methods. Let us begin with the
perturbative method:
\subsubsection*{Perturbative Approach}
This identity will be shown using the LSZ formalism for the
correlation functions. As in the ordinary commutative QED an
arbitrary diagram is to be considered which involves an arbitrary
number of fermion and photon propagators, that are connected by
vertices of two fermions and one gauge field. Let us insert a
photon line into this diagram. According to the Feynman rules of
noncommutative dipole U(1) gauge theory, the only possibility to
insert this photon to the diagram is either to attach it to an
open fermion line or to an fermion loop. Consider first a part of
the correlation function corresponding to the $j$-th fermion line
in the momentum space before inserting the photon line to it:
\begin{eqnarray}\label{S81}
 {\mathcal{M}}_{0j}(k)=\Biggr[\prod^{n}_{\ell=1}\frac{i}{\PS_{\ell}-m}
 \Big(-2g\gamma^{\mu_{\ell}}\sin\Big(
 \frac{q_{\ell}.L}{2}\Big)\Big)\Biggr]\frac{i}{\PS-m}.
\end{eqnarray}
Now let us attach our photon first between the $r$-th and the
$r+1$-th vertices.: \vskip1cm
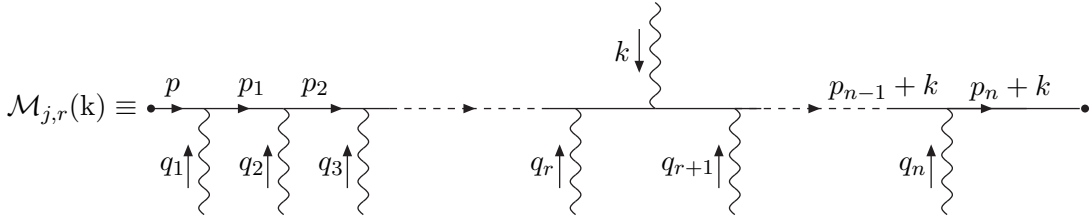
\begin{figure}[ht]
\begin{center}
\SetScale{1}
\begin{picture}(250,60)(0,0)
\Vertex(-40,20){1.5} \ArrowLine(-40,20)(-20,20)
\ArrowLine(-20,20)(10,20) \ArrowLine(10,20)(50,20)
\DashArrowLine(50,20)(110,20){3}
\Line(110,20)(150,20)\Photon(150,60)(150,20){2}{4}
\Photon(120,-20)(120,20){2}{4} \Photon(180,-20)(180,20){2}{4}
\DashArrowLine(180,20)(240,20){3} \Line(150,20)(190,20)
\Line(240,20)(310,20) \Photon(-20,-20)(-20,20){2}{4}
\Photon(10,-20)(10,20){2}{4} \Photon(40,-20)(40,20){2}{4}
\Photon(260,-20)(260,20){2}{4} \ArrowLine(260,20)(290,20)
\Vertex(312,20){1.5} \LongArrow(-26,-10)(-26,5)
\LongArrow(4,-10)(4,5) \LongArrow(34,-10)(34,5)
\LongArrow(114,-10)(114,5)
\LongArrow(174,-10)(174,5)\LongArrow(254,-10)(254,5)
\LongArrow(144,50)(144,35) \Text(138,42)[]{$k$}
\Text(-32,-2)[]{$q_{1}$} \Text(-2,-2)[]{$q_{2}$}
\Text(28,-2)[]{$q_{3}$} \Text(248,-2)[]{$q_{n}$}
\Text(-32,28)[]{$p$} \Text(-2,28)[]{$p_{1}$}
\Text(22,28)[]{$p_{2}$} \Text(108,-2)[]{$q_{r}$}
\Text(163,-2)[]{$q_{r+1}$}\Text(237,28)[]{$p_{n-1}+k$}
\Text(285,28)[]{$p_{n}+k$}
\Text(-70,20)[]{${\mathcal{M}}_{j,r}(\mbox{k})\equiv$}
\end{picture}
\end{center}
\caption{A photon with the momentum $k$ is attached to a part of
the correlation function corresponding to the $j$-th fermion line.
}
\end{figure}
 \vskip1cm\par\noindent
 Here, $p_{i}=p_{i-1}+q_{i}$ and the index
$r$ labels the new vertex, which is created by this attachment. We
arrive at:
\begin{eqnarray*}
 {\mathcal{M}}_{j,r}(k)=\epsilon_{\mu}{\mathcal{M}}^{\mu}_{j,r}(k),
\end{eqnarray*}
with
\begin{eqnarray}\label{S82}
\lefteqn{\hspace{-1cm}
 k_{\mu}{\mathcal{M}}^{\mu}_{j,r}(k)=}\nonumber\\
 &&\hspace{-1cm}=
 \frac{i}{\PS_{n}+\KS-m}
 \Big(-2g\gamma^{\mu_{n}}\sin\Big(\frac{q_{n}.L}{2}\Big)\Big)\cdots
 \frac{i}{\PS_{r}+\KS-m}
 \Big(-2g\KS\sin\Big(\frac{k.L}{2}\Big)\Big)\frac{i}{\PS_{r}-m}\cdots\frac{i}{\PS-m}.
\end{eqnarray}
Using now the identity $\KS=(\PS_{r}+\KS-m)-(\PS_{r}-m)$ we
obtain:
\begin{eqnarray}\label{S83}
\lefteqn{
 k_{\mu}{\mathcal{M}}^{\mu}_{ j,r}(k)=2g\sin
 \Big(\frac{k.L}{2}\Big)
 }\nonumber\\
 &&\times \Biggr[\prod^{n}_{\ell=1}\Big(\frac{i}{\PS_{\ell}
 +\theta(\ell-r-1)\KS-m}-\frac{i}{\PS_{\ell}
 +\theta(\ell-r)\KS-m}\Big)
 \Big(-2g\gamma^{\mu_{\ell}}\sin\Big(\frac{q_{\ell}.L}{2}\Big)\Big)\Biggr]
 \frac{i}{\PS-m},
\end{eqnarray}
where $\theta(n)$ is the step function defined by
\begin{displaymath}
\theta(n)\equiv\Biggr\{\begin{array}{cc} 1&n\geq 0\\
                                   0&n<0
\end{array}
, \hspace{1cm}\forall n\in Z.
\end{displaymath}
 There are indeed $n+1$
possibilities to attach the photon to the fermion line. Summing
over all these insertion points $r$, we arrive at:
\begin{eqnarray}\label{S84}
\lefteqn{
 \sum\limits_{r=0}^{n}k_{\mu}{\mathcal{M}}^{\mu}_{
 j,r}(k)=}\nonumber\\
 &&=\sum^{n}_{r=1}\Biggr[\prod^{n}_{\ell=1}\Big(\frac{i}{\PS_{\ell}+
 \theta(\ell-r-1)\KS-m}-\frac{i}{\PS_{\ell}+
 \theta(\ell-r)\KS-m}\Big)\Big(-2g\gamma^{\mu_{\ell}}
 \sin\big(\frac{q_{\ell}.L}{2}\big)\Big)\Biggr]
\nonumber\\
 &&\times \Big(2g\sin\big(\frac{k.L}{2}\big)\Big)\frac{i}{\PS-m}+\frac{i}{\PS_{n}+\KS-m}
 \Big(-2g\gamma^{\mu_{n}}\sin\big(
 \frac{q_{n}.L}{2}\big)\Big)\cdots
 \nonumber\\
 &&\times \cdots \frac{i}{\PS+\KS-m}\Big(-2g\KS
 \sin\big(\frac{k.L}{2}\big)\Big)\frac{i}{\PS-m}.
\end{eqnarray}
After an appropriate redefinition of the summation index in the
second term of the above expression and after a long but straight
forward calculation, we arrive at:
\begin{eqnarray}\label{S85}
 \sum\limits_{r=0}^{n}k_{\mu}{\mathcal{M}}^{\mu}_{
 j,r}(k)=2g\sin\big(\frac{k.L}{2}\big)
 \Biggr[\prod^{n}_{\ell=0}\Big(\frac{i}{\PS_{\ell}-m}-\frac{i}{\PS_{\ell}+\KS-m}\Big)
 \Big(-2g\gamma^{\mu_{\ell}}\sin\big(\frac{q_{\ell}.L}{2}\big)\Big)\Biggr].
\end{eqnarray}
As next, let us sum over all possible fermion lines, ({\it i.e.}
sum over all $j$'s). We arrive at the Ward-Takahashi identity for
the correlation functions, which  reads:
\begin{eqnarray}\label{S86}
 \lefteqn{k^{\mu}{\mathcal{M}}_{\mu}\big(k;\{p_{i},q_{i}\};L\big)=2g\sin\big(\frac{k.L}{2}\big)}
\nonumber\\
&&\times
 \sum_{j}\Big[{\mathcal{M}}_{0}\big(\{p_{i}\};q_{1},\cdots,q_{j}-k,\cdots,q_{n};L\big)-
 {\mathcal{M}}_{0}\big(p_{1},
 \cdots,p_{j}+k,\cdots,p_{n};\{q_{i}\};L\big)\Big].
\end{eqnarray}
Using now the standard LSZ formalism, and considering only the
on-shell external momenta $\{p_{i}\}$ and $\{q_{i}\}$, the r.h.s.
of the above equation vanishes by usual argumentation from the
commutative U(1) gauge theory. The WT identity for the $S$-matrix
elements is therefore given by:
\begin{eqnarray}\label{S87}
k^{\mu}{\mathcal{M}}_{\mu}\big(k;\{p_{i},q_{i}\};L\big)=0.
\end{eqnarray}
The same expression is indeed valid if the photon is inserted to a
fermion loop. This can be seen by putting $p_{0}=p_{n}$ in the Eq.
(\ref{S85}).
\subsubsection*{Path Integral Approach}
In this section, the WT identity of noncommutative dipole U(1)
gauge theory with matter fields in the adjoint representation will
be calculated using the non-perturbative path integral method.
Using the local gauge transformation of the matter fields from Eq.
(\ref{S20}), it can be shown that the inner product
$\bar{\psi}\star\psi$ is invariant:
\begin{eqnarray}\label{S88}
 \bar{\psi}\star\psi\longrightarrow \big(U\star(\bar{\psi}\star\psi)\big)\star
 U^{-1}=(U^{-1}U)(\bar{\psi}\star\psi)=\bar{\psi}\star\psi.
\end{eqnarray}
Note that both $U\in U(1)$ and the product $\bar{\psi}\star\psi$
are dipoleless. Let us now consider the following expression
\begin{eqnarray}\label{S89}
 F_{\{\mu_{i},\nu_{i}\}}\big(\{x_{i},x^{\prime}_{i}\}\big)=
 \int D\psi\  D\bar{\psi}\ DA\ e^{i\int{d^{4}x{\mathcal{L}}^{adj}_
 {QED}}}\ T\left(\prod^{n}_{i=1}\psi_{\mu_{i}}(x_{i})\bar{\psi}_{\nu_{i}}(x^{\prime}_{i})
 \right).
\end{eqnarray}
Under an infinitesimal gauge transformation of the matter fields
in the form given in the Eq. (\ref{S28}), we have:
\begin{eqnarray}\label{S90}
0&=&
 \int{D\bar{\psi}\ D\psi\  DA\ e^{i\int{d^{4}x{\mathcal{L}}^{adj}_{QED}}}} \Biggr[\Big(\prod^{n}_{i=1}\psi_{\mu_{i}}
 (x_{i})\bar{\psi}_{\nu_{i}}(x^{\prime}_{i})\Big)\int{d^{4}x\big(\bar{\psi}\gamma^{\mu}\star[\partial_{\mu}\alpha,
 \psi]_{\star}\big)(x)}
 \nonumber\\
 &&
 -\Big(\sum^{n}_{i=1}\psi_{\mu_{1}}(x_{1})\cdots[\alpha,\psi_{\mu_{i}}]_{\star}(x_{i})\cdots\psi_{\mu_{n}}(x_{n})
 \Big)\Big(\prod^{n}_{j=1}\bar{\psi}_{\nu_{j}}(x^{\prime}_{j})\Big)
 \nonumber\\
&&-\Big(\prod^{n}_{j=1}\psi_{\mu_{j}}(x_{j})\Big)\Big(\sum^{n}_{i=1}\bar{\psi}_{\nu_{1}}(x^{\prime}_{1})\cdots
 [\alpha,\bar{\psi}_{\nu_{i}}]_{\star}(x^{\prime}_{i})
 \cdots\bar{\psi}_{\nu_{n}}(x^{\prime}_{n})\Big)\Biggr].
\end{eqnarray}
Here, Eq. (\ref{S29}) and the invariance of the measure of the
integral under the infinitesimal transformation from Eq.
(\ref{S28}) are used. After a partial integration in the first
term on the r.h.s. of Eq. (\ref{S90}) and using the relation:
\begin{eqnarray}\label{S91}
 [\alpha,\psi_{\mu_{i}}]_{\star}(x_{i})=-\int{d^{4}x\psi_{\mu_{i}}(x_{i})
 \alpha(x)\Big(\delta(x-x_{i}-\frac{L}{2})
 -\delta(x-x_{i}+\frac{L}{2})\Big)},
\end{eqnarray}
and the corresponding relation for $\bar{\psi}$ in the second and
third terms of Eq. (\ref{S90}), we have:
\begin{eqnarray}\label{S92}
\lefteqn{
 ik_{\mu}\big<\Omega|T\widetilde{J}^{\mu}(k)\prod^{n}_{i=1}\widetilde{\psi}_{\mu_{i}}(\ell_{i})\widetilde{\bar{\psi}}_
 {\nu_{i}}(\ell^{\prime}_{i})|\Omega\big>=
 }\nonumber\\
 &&\times 2ig\sin(\frac{k\cdot L}{2})
 \Biggr[\big<\Omega|T\Big(\sum^{n}_{i=1}\widetilde{\psi}_{\mu_{1}}(\ell_{1})\cdots\widetilde{\psi}_{\mu_{i}}(\ell_{i}-k)
 \cdots\widetilde{\psi}_{\mu_{n}}(\ell_{n})\Big)\Big(\prod^{n}_{j=1}\widetilde{\bar{\psi}}_{\nu_{j}}(\ell^{\prime}_{j})
 \Big)|\Omega\big>\nonumber\\
 &&
\hspace{3cm}
 -\big<\Omega|T\Big(\prod^{n}_{j=1}\widetilde{\psi}_{\mu_{j}}(\ell_{j})\Big)\Big(\sum^{n}_{i=1}\widetilde{\bar{\psi}}_
 {\nu_{1}}(\ell^{\prime}_{1})\cdots\widetilde{\bar{\psi}}_{\nu_{i}}(\ell^{\prime}_{i}+k)\cdots\widetilde{\bar{\psi}}_
 {\nu_{n}}(\ell^{\prime}_{n})\Big)|\Omega\big>\Biggr].
\end{eqnarray}
This relation can be finally rewritten in the form given in Eq.
(\ref{S86}), that we obtained in the previous section using the
perturbative method. Eq. (\ref{S87}) can therefore be given for
the S-matrix elements, too. \section{Renormalization Constants and
the one-loop $\beta$-Function} \setcounter{equation}{0} In this
section the renormalization constants of the noncommutative dipole
QED with adjoint matter fields will be studied non-perturbatively
and the one-loop $\beta$-function of the noncommutative QED with
adjoint matters will be derived explicitly. Using the WT identity,
which is proved in the previous section, we will first show that
$Z_{1}=Z_{2}$ in all orders of perturbative expansion. We then
turn to $Z_{3}$ and argue that due to the gauge invariance arising
from WT identity, the vacuum polarization amplitude is in all
order of the perturbative expansion proportional to the usual
tensorial structure $(p_{\mu}p_{\nu}-\eta_{\mu\nu}p^{2})$, where
$p$ is the external photon momentum. Note that the same situation
occurs in the ordinary commutative QED but in no way in the Moyal
case.
\subsection{$Z_{1}$ and $Z_{2}$} Let us consider the vertex
function $\Gamma^{\mu}(p,q)$ in the momentum space with the
incoming external momentum $q$ for the photon and outgoing
momentum $p$ for the fermion [See Fig. 5]: \vskip0.5cm
\begin{figure}[ht]
\begin{center}
\begin{picture}(200,40)(0,0)
 \CCirc(-40,0){10}{Black}{Gray}
 \Photon(-75,0)(-50,0){2}{4}
 \Vertex(-75,0){1.5}
 \ArrowLine(-12,-30)(-34,-8)
 \Vertex(-12,-30){1.5}
 \Text(-16,-15)[]{$p$}
 \ArrowLine(-34,8)(-12,30)
 \Vertex(-12,30){1.5}
 \Text(2,15)[]{$p+q=p'$}
 \LongArrow(-65,-6)(-55,-6)
 \Text(-62,+8)[]{$q$}
 \Text(36,0)[]{$=$}
 \Text(72,0)[]{$2g\sin\big(\frac{q.L}{2}\big)\Biggr[$}
 \CCirc(120,0){10}{Black}{Gray}
 \ArrowLine(120,-30)(120,-10)
 \Vertex(120,-30){1.5}
 \Text(128,-20)[]{$p$}
 \ArrowLine(120,10)(120,30)
 \Vertex(120,30){1.5}
 \Text(128,20)[]{$p$}
 \Text(150,0)[]{$-$}
 \CCirc(177,0){10}{Black}{Gray}
 \ArrowLine(177,-30)(177,-10)
 \Vertex(177,-30){1.5}
 \Text(195,-20)[]{$p+q$}
 \ArrowLine(177,10)(177,30)
 \Vertex(177,30){1.5}
 \Text(195,20)[]{$p+q$}
 \Text(215,0)[]{$\Biggr]$}
\end{picture}
\end{center}
\vskip0.5cm \caption{}
\end{figure}
\vskip0.5cm\par\noindent Making use of the WT identity from Eq.
(\ref{S86}), we first obtain:
\begin{eqnarray}\label{S93}
 S(p+q)\big(-2igq_{\mu}\Gamma^{\mu}(p+q,p)\big)S(p)=
 2g\sin\Big(\frac{q.L}{2}\Big)\big(S(p)-S(p+q)\big),
\end{eqnarray}
where $S(p)$ is the exact fermion propagator.  Taking as next the
zero momentum limit $q\to 0$, the vertex function (\ref{S93}) can
be given by:
\begin{eqnarray}\label{S94}
\Gamma^{\mu}(p+q,p)=Z_{1}^{-1}\gamma^{\mu}\sin\Big(\frac{q.L}{2}\Big),
\end{eqnarray}
which indeed gives an appropriate definition for $Z_{1}$. Using,
as usual the LSZ formalism, the only relevant term in the exact
fermion propagator is given by:
\begin{eqnarray}\label{S95}
S(p)=\frac{iZ_{2}}{\PS-m}.
\end{eqnarray}
Replacing as next the vertex function and the fermion propagator
from Eq. (\ref{S94}) and (\ref{S95}) in the Eq. (\ref{S93}), we
obtain $Z_{1}=Z_{2}$, as expected. This result is obtained without
any use of perturbation theory, and is therefore exact for all
orders of perturbative expansion. \subsection{The General
Structure of Vacuum Polarization Tensor and $Z_{3}$} Let us
consider the full photon propagator $D_{\mu\nu}(p,L)$ with $p$ the
external photon momentum and $L$ the dipole length. As in the
ordinary QED $D_{\mu\nu}(p,L)$ is given as a series of 1PI vacuum
polarization tensor $i\Pi_{\mu\nu}(p,L)$, which includes all loop
corrections [see Fig. 6].
\begin{figure}[ht]
\begin{center}
\begin{picture}(200,20)(0,0)
\Text(-100,0)[]{$D_{\mu\nu}\equiv$} \Photon(-50,0)(-30,0){2}{3}
\CCirc(-20,0){10}{Black}{Gray} \Photon(-10,0)(10,0){2}{3}
\Text(-45,8)[]{$\mu$} \Text(5,6)[]{$\nu$} \Text(20,0)[]{$=$}
\Photon(30,0)(90,0){2}{6} \Text(35,8)[]{$\mu$}
\Text(85,6)[]{$\nu$} \Text(98,0)[]{$+$}
\Photon(105,0)(130,0){2}{3} \CCirc(140,0){10}{Black}{Gray}
\Text(140,0)[]{1PI} \Photon(150,0)(175,0){2}{3}
\Text(110,8)[]{$\mu$} \Text(170,6)[]{$\nu$} \Text(182,0)[]{$+$}
\Photon(190,0)(215,0){2}{3} \CCirc(225,0){10}{Black}{Gray}
\Text(225,0)[]{1PI} \Photon(235,0)(245,0){2}{2}
\CCirc(255,0){10}{Black}{Gray} \Text(255,0)[]{1PI}
\Photon(265,0)(290,0){2}{3} \Text(195,8)[]{$\mu$}
\Text(285,6)[]{$\nu$} \Text(305,0)[]{$+\cdots$}
\end{picture}
\end{center}
 with
\begin{center}
\begin{picture}(200,10)(0,0)
\Text(140,0)[]{$\ \ \ \equiv\ \ \ i\Pi_{\mu\nu}(p,L)$}
\Photon(35,0)(60,0){2}{3} \CCirc(70,0){10}{Black}{Gray}
\Text(70,0)[]{1PI} \Photon(80,0)(105,0){2}{3} \Text(40,8)[]{$\mu$}
\Text(100,6)[]{$\nu$}
\end{picture}
\end{center}
\caption{The full photon propagator can be written as a series of
one particle irreducible (1PI) diagrams.}
\end{figure}
\par\noindent
According to the WT identity $D_{\mu\nu}(p,L)$ must satisfy the
relation:
\begin{eqnarray}\label{S96}
p^{\mu}D_{\mu\nu}(p,L)=0=p^{\nu}D_{\mu\nu}(p,L).
\end{eqnarray}
The same relation must therefore be satisfied by each
$i\Pi_{\mu\nu}(p,L)$. The only two index object can be made from
$p_{\mu}$, $\eta_{\mu\nu}$ and $L_{\mu}$ which satisfies both
sides of the above equation is given by
\begin{eqnarray}\label{S97}
i\Pi_{\mu\nu}(p,L)=4i\sin^{2}\big(\frac{p\cdot
L}{2}\big)(p_{\mu}p_{\nu}-\eta_{\mu\nu}p^{2})\Pi(p,L).
\end{eqnarray}
Here we have separated the contribution of two vertices $V_{\mu}$
and $V_{\nu}$ [Eq. (\ref{S21})] as a factor
$-4\sin^{2}\frac{p\cdot L}{2}$ before $\Pi(p,L)$. Note that the
above tensor structure does not include $L$. Remember that in the
Moyal case no such simple tensor structure can be obtained for the
vacuum polarization tensor \cite{h15}.
\par
Now, summing up the 1PI diagrams from the r.h.s. of the Fig. 6, we
obtain:
\begin{eqnarray}\label{S98}
 D_{\mu\nu}(p,L)=\frac{-i\ \eta_{\mu\nu}}{p^{2}
 \big[1+4\sin^{2}(\frac{p\cdot L}{2})\ \Pi(p,L)\big]},
\end{eqnarray}
where a term proportional to $\frac{p_{\mu}p_{\nu}}{p^{4}}$ on the
r.h.s. of the above equation is neglected, because it does not
contribute in the $S$-matrix elements. Here $\Pi(p,L)$ includes
all radiative corrections.
\par
The general structure of $Z_{3}$ can now be defined using the  Eq.
(\ref{S98}) where the small momentum limit $p_{i}\to 0, i=1,2,3$
is to be considered
\begin{eqnarray}\label{S99}
 Z_{3}=\frac{1}{\Big[1+(p\cdot L)^{2}\Pi(0,L)\Big]}.
 \end{eqnarray}
The factor $(p\cdot L)^{2}$ which appears before $\Pi(0,L)$ in the
denominator of the above equation arises from the $p_{i}\to 0$
limit of $\sin^{2}(\frac{p\cdot L}{2})$ in the denominator of the
expression on the r.h.s of the Eq. (\ref{S98}). As it is known, in
the ordinary QED, where no $\sin^{2}(\frac{p\cdot L}{2})$ factor
appears in the denominator of $Z_{3}$, the renormalization
constant $Z_{3}$ is defined by the function $\Pi(p^{2})$ evaluated
at $p^{2}$ exactly equal to zero. It can be shown that in the
commutative case the function $\Pi(p^{2})$ coming from the vacuum
polarization tensor includes both $p$-dependent and
$p$-independent parts. In the limit $p_{i}\to 0$ the $p$-dependent
part can be neglected comparing to the $p$-independent part
because of its polynomial structure. This is the same as putting
$p_{i}$ exactly equal zero, from the begin on, in the original
commutative $\Pi(p^{2})$.
\par
In the noncommutative dipole Field Theory, however, $\Pi(p,L)$ is
{\it ab initio} multiplied by $\sin^{2}(\frac{p\cdot L}{2})$. This
means that no $p$-independent terms appear in the denominator of
$\Pi_{\mu\nu}(p,L)$ and $Z_{3}$. The factor $(p\cdot L)^{2}$
survives therefore the $p_{i}\to 0$ ($i=1,2,3,$) limit. Using now
the Eq. (\ref{S80}), $\delta Z_{3}$ can be given in all order of
perturbative expansion by:
\begin{eqnarray}\label{S100}
\delta Z_{3}=-(p\cdot L)^{2}\  \Pi(0,L),
\hspace{1cm}\mbox{with}\hspace{1cm}p_{i}\to 0.
\end{eqnarray}
\subsection{One-loop $\beta$-Function of Noncommutative Dipole QED}
In this section, we will first calculate the one-loop
$\beta$-function of noncommutative dipole QED with adjoint matters
explicitly. We will show that it is proportional to the same
factor $(p\cdot L)^{2}$, which appears also in $\delta Z_{3}$ from
Eq. (\ref{S100}). Using a semi-classical analysis we will then
explain the physical origin of this factor.
\par
Following the standard procedure from commutative QED, we
calculate the one-loop $\beta$-function, by separating the
Lagrangian of the theory in two parts, the renomalized part and
the counterterm part. The renomalized part is in terms of
renormalized fields $\psi_{r}, A_{\mu,r}$ and renormalized
coupling constant $g$. They are related to bare fields $\psi$ and
$A_{\mu}$ and bare coupling constant $g_{0}$ through standard
commutative relations:
\begin{eqnarray}\label{S102}
 \psi_{r}=Z_{2}^{-1/2}\psi,\hspace{1cm}
 A_{\mu,r}=Z_{3}^{-1/2}A_{\mu},\hspace{1cm}
 \mbox{and}\hspace{1cm} g_{0}=g\ Z_{3}^{-1/2}Z_{1}Z_{2}^{-1}=g\ Z_{3}^{-1/2}.
\end{eqnarray}
In the last expression, the WT identity from previous section is
used ($Z_{1}=Z_{2}$). All loop diagrams which arise from the
perturbative expansion using this Lagrangian, depend therefore on
renomalized coupling constant $g$, which  must be taken as a
function of the renomalization ({\it energy}) scale $\mu$. The
$\beta$-function of the theory is then defined by the variation of
$g$ with respect to $\mu$:
\begin{eqnarray}\label{S103}
\beta(g)=\mu\frac{\partial g(\mu)}{\partial\mu}.
\end{eqnarray}
According to the last expression in Eq. (\ref{S102}) the
renormalized coupling constant is proportional to $Z_{3}$. In the
one-loop order it is given by
\begin{eqnarray}\label{S104}
 Z_{3}=1+\delta Z_{3}=1+\frac{g^{2}(\mu)}{(4\pi)^{d/2}}\ \sin^{2}(\frac{\vec{p}\cdot
\vec{L}}{2})\int_{0}^{1}
 dx\ \big(8x(1-x)\big)
 \frac{\Gamma(\frac{\epsilon}{2})}{[-x(1-x)p^{2}]^{\epsilon/2}}
 \Biggr|_{(p_{0}=\mu, \vec{p}^{2}<<\mu^{2})}.
\end{eqnarray}
The factor $\sin(\frac{\vec{p}\cdot\vec{L}}{2})$ comes from the
vertex of two fermion and one gauge field from Eq. (\ref{S21}).
Taking $p_{0}=\mu$ and $\vec{p}^{2}<<\mu^{2}$ the renormalization
condition $p^{2}=p_{0}^{2}-\vec{p}^{2}=\mu^{2}$ is guaranteed.
Putting now $p=(p_{0}=\mu, \vec{p})$ with small
$\vec{p}^{2}<<\mu^{2}$ and $L=(0,\vec{L})$, the factor
$\sin(\frac{p\cdot L}{2})$ can be replaced by
$(\vec{p}\cdot\vec{L}/2)$. The one-loop $\beta$-function of the
noncommutative dipole QED is then by:
\begin{eqnarray}\label{S105}
\beta(g)=-\frac{g^{3}(\mu)}{12\pi^{2}}\ \alpha^{2}(\vec{p},
\vec{L}),
\end{eqnarray}
where we have introduced the function
$\alpha(\vec{p},\vec{L})\equiv \vec{p}\cdot \vec{L}$ with
$\vec{p}^{2}<<\mu^{2}$.  The minus sign of $\beta(g)$ indicates
that the noncommutative QED with adjoint matters is asymptotically
free \cite{h6}. According to this result the one-loop
$\beta$-function of the theory depends on the product
$\vec{p}\cdot\vec{L}\equiv |\vec{p}|\ |\vec{L}|\ \cos\vartheta$
with $\vartheta$ the relative angle between  the external momentum
$\vec{p}$ and the dipole length $\vec{L}$. This means that the
function $\beta(g)$ vanishes not only at $|\vec{L}|=0$, as
expected\footnote{For $|\vec{L}|=0$ noncommutative dipole QED with
adjoint matters turns out to be free.}, but also when either
$|\vec{p}|=0$ (forward scattering) or when $\theta=\frac{\pi}{2}$
or $\frac{3\pi}{2}$. The theory, however, remains always
asymptotically free, because the factor $\alpha^{2}(\vec{p}\cdot
\vec{L})$ in the one-loop $\beta$-function is always positive. The
appearance of the scalar product $(\vec{p}\cdot\vec{L})$ in the
argument of the sine is due to the broken Lorentz symmetry of the
noncommutative dipole QED, which has its origin in the appearance
of a fixed vector $\vec{L}$ with a definite length and direction.
\par
In the following we would like to give a physical interpretation
of the factor $\alpha(\vec{p}, \vec{L})$ in the one-loop
$\beta$-function from Eq. (\ref{S105}):
\par
As is well-known, in the QFT, in general,  the results of the
scattering amplitudes in the small momentum limit ($p_{i}\to 0,\
i=1,2,3$) must coincide with the classical results. In the Born
approximation the scattering amplitudes are proportional to the
Fourier transformed of the scattering potential energy. In a
theory with finite point like charges, the potential
$V_{e}(\vec{r})$ is a central Coulomb potential
$V_{e}(r)=\frac{g}{4\pi\epsilon_{0}r}$. Hence in the framework of
QED for the small momentum limit, the scattering amplitudes of the
electrons are proportional to
$\widetilde{V}_{e}(p)=\frac{g^{2}}{p^{2}}$ in the momentum space.
\par
In the noncommutative dipole QED with matter fields in the adjoint
representation, however, no point like charged particles are
present. The theory includes only finite multipoles. In the lowest
order of the dipole length $L$, the scattering potential energy is
therefore the energy between two dipoles. According to the Born
approximation, the classical scattering amplitudes are
proportional to the Fourier transformed of the potential energy of
{\it two dipoles} $g_{0}L_{1}$ and $g_{0}L_{2}$ at large
distances. For $L_{1}=L_{2}\equiv L$ this classical potential
energy in the momentum space is given by [see Appendix A for a
derivation]:
\begin{eqnarray}\label{S106}
 \widetilde{V}_{L}(p)\Biggr|_{\mbox{\small{classical}}}=-\frac{\bar{g}_{0}^{2}}
 {p^{2}},\hspace{1cm}\mbox{with}\hspace{1cm}\bar{g}_{0}\equiv
 g_{0}\alpha(\vec{p}, \vec{L}),
\end{eqnarray}
and $\alpha(\vec{p},\vec{L})\equiv \vec{p}\cdot\vec{L}$. Hence
comparing to the potential of a point like charged particle
$\widetilde{V}_{e}(p)$, the bare coupling constant must be
replaced by $\bar{g}_{0}$, and is therefore proportional to the
factor  $\alpha(\vec{p},\vec{L})$. This replacement is only true
for a process where two photons with the momentum $p$ are involved
as in Fig. 7
\begin{figure}[ht]
\begin{center}
\begin{picture}(200,20)(0,0)
\ArrowLine(-40,-20)(-20,0) \Vertex(-20,0){1}
\ArrowLine(-20,0)(-40,20) \Photon(-20,0)(0,0){2}{3}
\CCirc(10,0){10}{Black}{Gray} \Photon(20,0)(40,0){2}{3}
\Vertex(40,0){1} \ArrowLine(40,0)(60,20) \ArrowLine(60,-20)(40,0)
\LongArrow(70,0)(110,0) \Text(90,-8)[]{$p_{i}\to 0$}
\ArrowLine(140,0)(120,20) \ArrowLine(120,-20)(140,0)
\Vertex(140,0){1} \Photon(140,0)(200,0){2}{6}
\ArrowLine(220,-20)(200,0) \ArrowLine(200,0)(220,20)
\Vertex(200,0){1}
\end{picture}
\end{center}
\vspace{0.5cm} \caption{}
\end{figure}
\vspace{.5cm}
\par\noindent
 The new renormalized coupling constant can then be
given by $\bar{g}\equiv \alpha(\vec{p}, \vec{L})g=\bar{g}_{0}\
Z_{3}^{1/2}$. This is exactly the same factor which appears in the
expression (\ref{S104}) for $Z_{3}$. Going now through the same
procedure as described above the one-loop $\beta$-function of
noncommutative dipole QED turns out to be given by the same Eq.
(\ref{S105}).
\section{Form Factors and One-Loop Anomalous Magnetic
Moment}\setcounter{equation}{0} In this section we will study the
form factors of noncommutative dipole QED with adjoint matter
fields and derive explicitly the anomalous magnetic moment of this
theory.
\par
The general form of the vertex function $\Gamma^{\mu}(p',p)$ of
noncommutative QED is given by:
\begin{eqnarray}\label{S107}
 \Gamma^{\mu}(p^{\prime},p)=
 \Big[\gamma^{\mu}A_{1}+\frac{(p+p^{\prime})^{\mu}}{m}A_{2}+\frac{(p^{\prime}-p)^{\mu}}{m}A_{3}
 +mL^{\mu}A_{4}\Big]\sin\Big(\frac{q\cdot L}{2}\Big),
\end{eqnarray}
where  according to the diagram on the l.h.s. of the Fig. 2, $p'$
and $p$ are the momenta of the external fermions and $q$ is the
photon momentum. The functions $A_{i}, i=1,2,3,4$ depend  in
general on
$\frac{q^{2}}{m^{2}},p.L,p^{\prime}.L,\frac{\PS}{m},\frac{\PS^{\prime}}{m},m\LLS$
which appear only in polynomial structures. Here $L$ is the dipole
length of the theory and for our physical purposes is a small
parameter. The last term on the r.h.s. of Eq. (\ref{S107}) can
therefore be neglected.  Besides the parameter $m\LLS$ does not
appear in the functions $A_{i}, i=1,2,3$ anymore. Following the
standard arguments form commutative QED and using the WT identity
$q_{\mu}\Gamma^{\mu}(p',p)=0$, we arrive at:
\begin{eqnarray}\label{S108}
 \Gamma^{\mu}(p^{\prime},p)=\Big[\gamma^{\mu}F_{1}(q^{2},p\cdot L,p^{\prime}\cdot L)
 +\frac{i\sigma^{\mu\nu}q_{\nu}}{2m}F_{2}
 (q^{2},p\cdot L,p^{\prime}\cdot L)\Big]\sin\Big(\frac{q\cdot
 L}{2}\Big),
\end{eqnarray}
where $\sigma^{\mu\nu}\equiv
\frac{i}{2}\big[\gamma^{\mu},\gamma^{\nu}\big]$ and $F_{i}, i=1,2$
are the form factors of the theory. Here, in comparison to the
commutative QED, they have new physical origins. Remember that in
the noncommutative dipole QED, no point like charged particles are
present. The theory includes only {\it multipoles}. In a
semi-classical approximation, the form factors are therefore
defined by the interaction of a finite dipole moment $g\vec{L}$
with external electric and magnetic fields. This is in contrast to
commutative QED, where the interaction of point like charged
particles with external fields defines the form factors.
\par
To compute $F_{1}(q^{2},p\cdot L,p^{\prime}\cdot L)$ let us take,
as in the commutative case, a classical external potential
$A^{\mu}_{cl}(x)=\big(\phi(\vec{x}),\vec{0}\big)$, where
$\phi(\vec{x})$ is an arbitrary electrical potential.  In the
momentum space, it can be given by $\widetilde{A}^{\mu}_{cl}(q)=
\big(2\pi\delta(q^{0})\widetilde{\phi}(\vec{q}),\vec{0}\big)$.
Using $\widetilde{A}^{\mu}_{cl}(q)$, the scattering amplitude of a
dipole moment can be calculated and reads:
\begin{eqnarray}\label{S109}
 i{\mathcal{M}}=
 ig\ \bar{u}(p^{\prime})\Gamma^{0}(p^{\prime},p)u(p)\ \widetilde{\phi}(\vec{q}).
\end{eqnarray}
Now going back to the Eq. (\ref{S108}) and taking the small
momentum limit $q_{i}\to 0, i=1,2,3$, we arrive at:
\begin{eqnarray}\label{S110}
\Gamma^{0}(p^{\prime},p)=2g
 \gamma^{0}\ \frac{\vec{q}\cdot\vec{L}}{2}\ F_{1}(0,L).
\end{eqnarray}
Putting now this expression on the r.h.s. of the Eq. (\ref{S109}),
we obtain:
\begin{eqnarray}\label{S111}
 i{\mathcal{M}}=-iF_{1}(0,L)\ (g\vec{L}\cdot\vec{q})\ \widetilde{\phi}(\vec{q})\ \big(2m\xi^{\prime
 \dag}\xi\big),
\end{eqnarray}
where $\xi$ and $\xi^{\dagger}$ are two component spinors. Going
back to the position space and comparing this relation with the
classical potential energy of an external electric field $\vec{E}$
acting on a dipole moment $g\vec{L}$
\begin{eqnarray*}
V(\vec{x})=-gF_{1}(0,L)\ \vec{L}\cdot\vec{E},
\end{eqnarray*}
we have
\begin{eqnarray}\label{S112}
F_{1}(0,L)=1.
\end{eqnarray}
This is in contrast with the commutative QED where, for the limit
of small momenta, the scattering amplitude is to be compared with
the Fourier transform of the classical potential energy of a {\it
point like charged particle} in the external field
$\phi(\vec{x})$.
\par
Taking now $A^{\mu}_{cl}(x)=\big(0,\vec{A}_{cl}(\vec{x})\big)$ as
an arbitrary magnetic potential with
$\widetilde{A}^{\mu}_{cl}(q)=\big(0,2\pi\delta(q^{0})\widetilde{A}^{i}_{cl}(\vec{q})\big)$
in the momentum space, the scattering amplitude of a dipole moment
can be first given by:
\begin{eqnarray}\label{S113}
 i{\mathcal{M}}=2ig\frac{\vec{q}\cdot\vec{L}}{2}\bar{u}(p^{\prime})\Big[\gamma^{i}F_{1}(0,L)+
 \frac{i\sigma^{i\nu}q_{\nu}}{2m}F_{2}(0,L)\Big]u(p)A^{i}_{cl}(\vec{q}).
\end{eqnarray}
Using now the standard relation:
\begin{eqnarray}\label{S114}
 \bar{u}(p^{\prime})\Big[\gamma^{i}F_{1}(0,L)+\frac{i\sigma^{i\nu}q_{\nu}}{2m}F_{2}(0,L)\Big]u(p)
 =2m\xi^{\prime\dag}
 \Bigg\{-\frac{i}{2m}\varepsilon^{ijk}q^{j}\sigma^{k}\bigg[F_{1}(0,L)+F_{2}(0,L)\bigg]\Bigg\}
 \xi,
\end{eqnarray}
with the Pauli matrices $\sigma^{k}, k=1,2,3$, we arrive at:
\begin{eqnarray}\label{S115}
 i{\mathcal{M}}=-(g\vec{L}\cdot\vec{q})\ (2m)\ \xi^{\prime\dag}\Big(\frac{\sigma^{k}}{2}\Big)
 \xi\  \big[\varepsilon^{kji}q^{j} \widetilde{A}^{i}_{cl}(\vec{q})\big]\
 \bigg[1+F_{2}(0,L)\bigg],
\end{eqnarray}
where we have used the previous result from Eq. (\ref{S112}).
Going again back to the position space and comparing this
relation with the classical potential energy of an external
magnetic field $\vec{B}$ acting on a {\it dipole moment}
$g\vec{L}$:
\begin{eqnarray*}
V(\vec{x})=-<\vec{\mu}>\cdot\Big(\frac{\vec{L}}{|\vec{L}|}.
\vec{\nabla}\Big)\vec{B}(\vec{x}),
\end{eqnarray*}
where the magnetic dipole moment $\vec{\mu}\equiv
\mbox{\emph{g}}_{\mbox{\small\emph{land\'{e}}}}
\Big(\frac{g|\vec{L}|}{2m}\Big)\vec{S}$ is introduced, we arrive
at:
\begin{eqnarray}\label{S115x}
 <\vec{\mu}>=\frac{g|\vec{L}|}{m}\ \bigg[F_{1}(0,L)+F_{2}(0,L)\bigg]\
 \xi^{\prime\dag}\vec{S}\xi,
\end{eqnarray}
with $\vec{S}=\vec{\sigma}/2$. The Land\'{e} factor is therefore
given by:
\emph{g}$_{\mbox{\small\emph{land\'{e}}}}=2[1+F_{2}(0,L)]$, where
$2F_{2}(0,L)$ is the anomalous part of this factor. It is now
possible to find the one-loop contribution to $F_{2}(0,L)$ from
the one-loop Feynman diagram corresponding to the vertex function
(\ref{S77}). We obtain
\begin{eqnarray}\label{S116}
F_{2}(0,L)&=&\frac{g^{2}}{\pi^{2}}m^{2}\vec{L}^{2}\int_{0}^{1}d\alpha\int_{0}^{1-\alpha}d\beta
\int_{0}^{1-\alpha-\beta}d\gamma\
\delta(\alpha+\beta+\gamma-1)\frac{\alpha(\beta+\gamma)}{(\alpha+\beta+\gamma)^{4}}\exp\Big(\frac{(\beta+\gamma)^{2}}
{\alpha+\beta+\gamma}\Big)\nonumber\\
&=&\frac{g^{2}}{2\pi^{2}}m^{2}\vec{L}^{2}.
\end{eqnarray}
The anomalous magnetic moment is therefore proportional to
$\vec{L}^{2}$ and vanishes by taking the limit $L\to 0$, as
expected from a free Field Theory.
\section{Conclusion}
In this paper, a detailed study of the noncommutative dipole QED
with matter fields in the adjoint representation is presented.
After introducing the action of the theory in Sect. 3,  the
Noether currents and the corresponding conserved charges are
derived using the noncommutative version of the Noether procedure.
In analogy to the noncommutative Moyal case \cite{h11},
noncommutative dipole QED possesses three different global vector
and axial vector currents.  In Sect. 4, the axial anomaly for all
these currents are calculated in two and four dimensions using the
point split and dimensional regularization methods. In \cite{h12},
the axial anomaly was calculated for only one of the three global
currents of the theory using the Fujikawa's path integral method
\cite{h14}. Our result coincides with the result presented in this
paper. We have further shown, that in two dimensions the axial
anomalies corresponding to all three currents of the theory vanish
after integrating over one space component. The axial charges
corresponding to these currents are therefore conserved. In four
dimensions, however, the axial charges corresponding to two
currents $J'_{\mu(5)}$ and $J''_{\mu(5)}$ from Eq. (\ref{S33}) are
anomalous, whereas the axial charge of $J_{\mu(5)}$ from Eq.
(\ref{S32}) is still conserved.
\par
In Sect. 5, the fermion- and photon-self energy and the vertex
function are calculated up to one-loop order. Comparing the
one-loop Feynman integrals of the fermion self-energy and the
vertex function, which include both planar and nonplanar parts, we
have shown that $Z_{1}=Z_{2}$, where $Z_{i}, i=1,2$ are the
standard renormalization constants. Further the vacuum
polarization tensor has the usual tensorial structure
$(\eta_{\mu\nu}p^{2}-p_{\mu}p_{\nu})$. These properties are also
valid for all higher orders of perturbative expansion. This could
be shown in a detailed analysis of the Ward-Takahashi identity of
the noncommutative QED with adjoint matter fields using two
different methods in Sect. 6. It shall be noticed, that the
noncommutative Moyal case has none of these two properties
\cite{h15}. This is mainly so, because in the Moyal
noncommutativity, even in one-loop order, additional diagrams
appear, which arise from additional three and four gauge vertices.
These vertices are absent in the noncommutative dipole QED.
\par
The general structure of the renormalization constant $Z_{3}$ is
presented in Sect. 7, and the one-loop $\beta$-function of the
theory is then calculated explicitly. We have found that comparing
to the commutative QED, the noncommutative dipole QED with adjoint
matters is asymptotically free. Further the $\beta$-function of
the theory is proportional to a factor $\alpha(\vec{p}, \vec{
L})\equiv (\vec{p}\cdot \vec{L})$ with small momentum $\vec{p}$.
In fact, it could be shown that $\vec{p}^{2}<<\mu^{2}$, where
$\mu$ is the renormalization scale parameter. As in the
semi-classical approximation of commutative QED, we expect that
for the small momenta $\vec{p}$, the scattering amplitudes must
coincide with the classical results of scattering processes.
According to the Born approximation, the scattering amplitude must
be proportional to the Fourier transformed of the potential
energy. Since in a noncommutative dipole theory with adjoint
matters, point like charged particles are absent, the potential
energy can only be defined between multipoles. In the lowest order
of multipole expansion, we calculated the potential energy between
two dipoles [see Appendix A], and found out that the bare coupling
constant of the noncomutative dipole QED must indeed be modified
by the factor $\alpha(\vec{p}, \vec{L})$. This is the same factor
which appears in the one-loop $\beta$-function of the theory.
\par
Since the one-loop $\beta$-function of the theory depends on the
product $\vec{p}\cdot\vec{L}\equiv |\vec{p}|\ |\vec{L}|\
\cos(\vartheta)$ with $\vartheta$ the relative angle between the
momentum $\vec{p}$ and the dipole length $\vec{L}$, it vanishes
not only at $|\vec{L}|=0$, as expected from a free theory, but
also when either $|\vec{p}|=0$, {\it i.e.} in a forward scattering
or when $\theta=\frac{\pi}{2}$ or $\frac{3\pi}{2}$. The theory,
however, remains always asymptotically free, because the factor
$\alpha^{2}(\vec{p}\cdot \vec{L})$ in the one-loop
$\beta$-function is always positive. Hence the value of the
one-loop $\beta$-function depends on the direction of the dipole
length $\vec{L}$. This is mainly because introducing a constant
vector $\vec{L}$ with a definite length and direction in the
theory breaks the Lorentz symmetry of the theory.
\par
In Sect. 8, the form factors of the theory are defined using a
semi-classical approximation. In contrary to the commutative QED,
where the form factors can be defined by studying the effect of
external electric and magnetic fields on point-like charged
particles,  the form factors in the noncommutative QED are to be
defined by the Fourier transformed of the potential energy between
the {\it dipole} and external electric and magnetic fields. The
anomalous magnetic moment is then calculated up to one-loop order.
We have shown, that it is proportional to $(m\vec{L})^{2}$ and
vanishes by taking $L\to 0$.
\section{Acknowledgement}
Both authors thank F. Ardalan and H. Arfaei for useful
discussions.

\begin{appendix}
\setcounter{equation}{0}
\section{Classical Potential of Two Electric Dipole Moments}
Let us consider two electric dipole moments $g\vec{L}_{1}$ and
$g\vec{L_{2}}$. The potential energy of the second dipole arising
from the first one is given by:
\begin{eqnarray}\label{A1}
 V_{L}(\vec{r})=g\vec{L}_{2}\cdot\vec{\nabla}\phi(\vec{r})\hspace{1cm}\mbox{with}\hspace{1cm}
 \phi(\vec{r})=\frac{g}{4\pi}\frac{\vec{L}_{1}\cdot\vec{r}}{r^{3}},
\end{eqnarray}
where $r$ is the distance between the center of two dipoles. In
this section we would like to calculate the Fourier transformed of
this potential energy explicitly. In the momentum space we have
\begin{eqnarray}\label{A2}
 \widetilde{V}(\vec{p})=\int{d^{3}re^{-i\vec{p}\cdot\vec{r}}V(\vec{r})}.
\end{eqnarray}
Taking  $\vec{p}$ along the $z$-axis and putting  Eq. (\ref{A1})
in Eq. (\ref{A2}), we arrive at:
\begin{eqnarray}\label{A3}
 \widetilde{V}(\vec{p})=\frac{g^{2}\vec{L}_{2}}{4\pi}\cdot
 \Big[\int{d^{3}r\ \vec{\nabla}\big(
 \frac{\vec{L}_{1}\cdot\vec{r}}{r^{3}}e^{-i\vec{p}\cdot\vec{r}}\big)}
 +i\vec{p}\int{d^{3}r\frac{\vec{L}_{1}\cdot\vec{r}}{r^{3}}e^{-i\vec
 {p}\cdot\vec{r}}}\Big],
\end{eqnarray}
where an integration by part is performed. Using the well-known
vector algebra identities, the first term is given by
\begin{eqnarray}\label{A4}
 g\int{d^{3}r\ \vec{\nabla}\big(\frac{\vec{L}_{1}\cdot\vec{r}}{r^{3}}
 e^{-i\vec{p}\cdot\vec{r}}\big)}=g
 \oint_{S}{d\vec{s}}\ \frac{\vec{L}_{1}\cdot
 \vec{r}}{r^{3}}\ e^{-i\vec{p}\cdot\vec{r}}=
 g\lim_{r\to\infty}\int{d\Omega
 \ \hat{r}\ \frac{\vec{L}_{1}\cdot\vec{r}}{r}\ e^{-ipr\cos\theta}}.
\end{eqnarray}
This term vanishes due to the factor $e^{-ipr\cos\theta}$ and the
symmetric integration interval of $\cos\theta\in [-1,1]$. Now let
us consider the second term on the r.h.s of the Eq. (\ref{A3}). In
the spherical coordinates, we obtain
\begin{eqnarray}\label{A5}
 \widetilde{V}(\vec{p})=\frac{ig^{2}}{2}(\vec{L}_{2}\cdot\vec{p})\ (\vec{L}_{1}\cdot \hat{z})
 \int{\frac{dr}{-ip}\frac{\partial}{\partial r}\int_{-1}^{1}{d\cos\theta}e^{-pr\cos\theta}}
 =-\frac{g^{2}(\vec{L}_{2}
 \cdot\vec{p})(\vec{L}_{1}\cdot\hat{z})}{p^{2}}\lim_{r\longrightarrow\
 0}\frac{\sin(pr)}{r}.
\end{eqnarray}
We finally arrive at:
\begin{eqnarray}\label{A6}
 \widetilde{V}(\vec{p})=-\frac{g^{2}(\vec{L}_{1}\cdot\vec{p})(\vec{L}_{2}\cdot
 \vec{p})}{p^{2}}.
\end{eqnarray}
This result is used in the Eq. (\ref{S106}).
\end{appendix}

\end{document}